\documentclass[preprint,5p,twocolumn,10pt]{elsarticle}

\usepackage[pdftex,hypertexnames=false,linktocpage=true]{hyperref}
\hypersetup{colorlinks=true,linkcolor=red,anchorcolor=darkgreen,citecolor=green!50!blue,filecolor=darkgreen,urlcolor=green,bookmarksnumbered=true,pdfview=FitB}

\usepackage{lipsum}
\usepackage{mathtools,cuted}

\usepackage{amsmath}

\usepackage{graphicx}

\usepackage{array}
\usepackage{booktabs}
\usepackage{arydshln}
\usepackage{multirow}

\usepackage{amsmath,amssymb,mathtools} 
\ifpdftex
    \usepackage{bbm}

\else
    \usepackage[bold-style=ISO]{unicode-math}
\fi

\usepackage{siunitx}  
\makeatletter
\newcommand{\vast}[1]{\bBigg@{#1}}
\makeatother

\newcommand{\intd}[3][]{\ifthenelse{\isempty{#3}}{\mathrm{d}^{#1} #2}{\frac{\mathrm{d}^{#1} #2}{#3}}\;}

\usepackage{stackengine}

\newcommand{\cond}{\; | \;}
\newcommand{\setcond}[2]{\ifthenelse{\isempty{#2}}{\{#1\}}{\{#1\cond{}#2\}}}

\usepackage{braket}
\usepackage{cancel}
    \def\slashed#1{\cancel{#1}}

\usepackage{braket}

\graphicspath{{../../plots/}}

\begin{document}

\begin{frontmatter}
\title{Transverse structure of the pion beyond leading twist with basis light-front quantization}

\author[imp,ucas,keylab]{Zhimin Zhu}
\ead{zhuzhimin@impcas.ac.cn}

\author[imp,ucas,keylab]{Zhi Hu\corref{cor1}}
\ead{huzhi@impcas.ac.cn}

\author[imp,ucas,keylab]{Jiangshan Lan\corref{cor1}}
\ead{jiangshanlan@impcas.ac.cn}

\author[imp,ucas,keylab]{Chandan Mondal\corref{cor1}}
\ead{mondal@impcas.ac.cn}

\author[imp,ucas,keylab]{Xingbo Zhao}
\ead{xbzhao@impcas.ac.cn}

\author[iowa]{James P. Vary}
\ead{jvary@iastate.edu}

\author[]{\\\vspace{0.2cm}(BLFQ Collaboration)}

\address[imp]{Institute of Modern Physics, Chinese Academy of Sciences, Lanzhou, Gansu, 730000, China}
\address[ucas]{School of Nuclear Physics, University of Chinese Academy of Sciences, Beijing, 100049, China}
\address[keylab]{CAS Key Laboratory of High Precision Nuclear Spectroscopy, Institute of Modern Physics, Chinese Academy of Sciences, Lanzhou 730000, China}
\address[iowa]{Department of Physics and Astronomy, Iowa State University, Ames, IA 50011, USA}

\cortext[cor1]{Corresponding author}

\begin{abstract}
  We investigate the twist-$3$ transverse-momentum-dependent parton distribution functions (TMDs) of the pion with basis light-front quantization. The twist-3 TMDs are not independent and can be decomposed into twist-$2$ and genuine twist-$3$ terms from the equations of motion (EOM). We compute the TMDs from the resulting light-front wave functions obtained by diagonalizing the light-front QCD Hamiltonian, determined for pion's constituent quark-antiquark and quark-antiquark-gluon Fock sectors, with three-dimensional confinement. We also obtain the twist-3 parton distribution functions (PDFs) and show that they preserve the sum rule, which affirms the robustness of our approach. This is the first time that theoretical predictions are made for subleading twist structures of the pion containing interference terms between two light-front Fock sectors.
\end{abstract}
\end{frontmatter}

\section{Introduction}
Understanding the structure of hadrons is one of the main goals of modern particle and nuclear physics~\cite{Accardi:2012qut,Bacchetta:2006tn,Belitsky:2002sm,Lai:2010vv,Pumplin:2002vw,Diehl:2003ny}. However, due to our insufficient knowledge of QCD color confinement and chiral symmetry breaking, it is difficult to obtain information on the internal structure of hadrons directly from the fundamental Lagrangian. On the other hand, with the development of the factorization theorem~\cite{Collins:2011zzd,Diehl:2011yj,Collins:1996fb,Diehl:2003ny,Rogers:2015sqa,Sterman:1995fz,Collins:1985ue,Collins:1987pm,Collins:1998rz}, one can decompose the cross sections of different experiments into short-distance perturbative and large-distance nonperturbative parts. The nonperturbative segment contains the structural information of the hadron~\cite{Collins:1987pm,Chernyak:1983ej,Muller:1994ses,Lampe:1998eu,Shuryak:1980tp}. For instance, one can extract the collinear parton distribution functions (PDFs) from deep inelastic scattering (DIS) processes~\cite{Hen:2016kwk,NNPDF:2017mvq}. The PDFs, describing the distribution of longitudinal momentum carried by the constituents inside hadrons~\cite{Jaffe:1983hp,Collins:2011zzd}, provide us with a one-dimensional (1D) longitudinal picture of hadrons. 
Therefore, the PDFs represent an initial 1D stage for comparing theory with experiment in the very challenging nonperturbative regime. This is our initial goal for addressing the structure of the pion.

Beyond the 1D structure, we gain insights into the three-dimensional (3D) structure of hadrons through the generalized parton distributions (GPDs)~\cite{Ji:1998pc, Polyakov:1999gs, Goeke:2001tz, Diehl:2003ny, Belitsky:2005qn} and the transverse-momentum-dependent parton distribution functions (TMDs)~\cite{Collins:2011ca, Angeles-Martinez:2015sea,Diehl:2015uka,Bacchetta:2016ccz}. The GPDs provide us with essential information about the distribution in position space and the orbital motion of partons inside hadrons~\cite{Burkardt:2000za,Burkardt:2002hr,Diehl:2003ny,Liu:2022fvl}. The GPDs are experimentally accessible through the deeply virtual Compton scattering (DVCS) or the deeply virtual meson production (DVMP) \cite{Belitsky:2005qn,Diehl:2003ny,Ji:1996nm,Favart:2015umi,Boffi:2007yc}. The TMDs provide information on the distribution of the transverse momentum in addition to the longitudinal momentum fraction \cite{Collins:2011zzd}. The TMDs, especially of quarks, can be extracted from the cross section associated with a specific azimuthal angle in the semi-inclusive deep inelastic scattering (SIDIS) processes or the Drell-Yan processes \cite{Collins:2011zzd,Boer:1997nt,Brodsky:2002rv,Brodsky:2002cx}. These two classes of distribution functions allow us to draw a 3D picture of hadrons.

Higher-twist parton distribution functions give access to a wealth of information about the partonic structure of hadrons~\cite{Jaffe:1983hp,Mulders:1995dh,Goeke:2005hb} such as describing multiparton correlations inside the hadrons that correspond to the interference between scattering from a single quark and from a coherent quark-gluon pair~\cite{Jaffe:1991ra,Jaffe:1991kp,Efremov:2002qh,Burkardt:2008ps}. They help us to understand the quark-gluon dynamics going beyond the probabilistic interpretation, which applies to the parton distribution functions at the leading-twist.
Twist-three PDFs and TMDs contribute to various observables in inclusive and semi-inclusive DIS, respectively.
Due to the existence of suppression factors, the twist-three contributions  are often much smaller than the twist-two.
However, in the kinematics of fixed target experiments, the twist-three structure functions are not small. 
One of the major goals of the future experimental program at JLab$12$ is to determine the different higher-twist spin-azimuthal asymmetries in SIDIS~\cite{E12017,E12-06-112,E12-06-112B}. 
Meanwhile, the upcoming Electron-Ion-Colliders (EICs) ~\cite{AbdulKhalek:2021gbh,AbdulKhalek:2022hcn,Amoroso:2022eow,Anderle:2021wcy} would access different kinematical regions~\cite{Boer:2011fh,Accardi:2012qut}. 
In this context, twist-three TMDs have been investigated in various QCD inspired models, for example, diquark spectator models~\cite{Jakob:1997wg,Lu:2012gu,Mao:2013waa,Mao:2014aoa}, the MIT bag model~\cite{Jaffe:1991ra,Signal:1996ct,Avakian:2010br,Lorce:2014hxa}, chiral quark soliton models~\cite{Schweitzer:2003uy,Wakamatsu:2007nc,Wakamatsu:2003uu,Ohnishi:2003mf,Cebulla:2007ej}, instanton models of QCD vacuum~\cite{Balla:1997hf,Dressler:1999hc}, constituent parton model on the light front~\cite{Lorce:2016ugb,Pasquini:2018oyz}, and perturbative light-front Hamiltonian approaches with a quark target~\cite{Burkardt:2001iy,Kundu:2001pk,Mukherjee:2009uy,Accardi:2009au}. 
Recently, the unpolarized twist-2 and twist-3, T-even quark TMDs in the pion have been evaluated by using the homogeneous Bethe-Salpeter integral equation in Ref.~\cite{Ydrefors:2023src}. 
In addition, phenomenological extractions of the pion's quark TMDs from  available cross section data of pion-nucleus induced Drell–Yan processes have been reported in Refs.~\cite{Cerutti:2022lmb,Barry:2023qqh}.

In this work, we present the twist-three TMDs of the pion obtained with basis light-front quantization (BLFQ), which provides a nonperturbative framework to solve relativistic many-body bound state problems in quantum field theories~\cite{Vary:2009gt,Zhao:2014xaa,Wiecki:2014ola,Li:2015zda,Jia:2018ary,Qian:2020utg,Tang:2019gvn,Mondal:2019jdg,Xu:2021wwj,Liu:2022fvl,Nair:2022evk,Lan:2021wok,Adhikari:2021jrh,Mondal:2021czk,Hu:2020arv,Lan:2019img,Lan:2019rba,Lan:2019vui,Xu:2022abw,Peng:2022lte}. Previously, this approach has been successfully applied to investigate the leading-twist TMDs of the physical spin-$\frac{1}{2}$ and spin-$1$ particles in QED, i.e., electron~\cite{Hu:2020arv} and photon~\cite{Nair:2022evk}, respectively, as well as for the proton~\cite{Hu:2022ctr}. 
We emphasize that we compute the pion TMDs at the model scale and do not consider the TMD evolution~\cite{Collins:2011zzd,Collins:1981va,Catani:2000vq,Bozzi:2005wk,Bozzi:2008bb}. 
Our calculation of TMDs should be physically interpreted as its nonperturbative, initial scale components. In other words, our model computation can provide a useful starting point for the TMDs at a specific scale, but the TMD evolution must be performed to obtain a complete distribution at a different scale. However, this is not the focus of the present work.


With the framework of BLFQ~\cite{Vary:2009gt}, we adopt the light-front QCD Hamiltonian~\cite{Lan:2021wok,Brodsky:1997de} and solve for its mass eigenvalues and eigenstates. With quark ($q$), antiquark ($\bar{q}$), and gluon ($g$) being the explicit degrees of freedom for the strong interaction, our Hamiltonian includes light-front QCD interactions relevant to constituent $|q\bar{q}\rangle$ and $|q\bar{q}g\rangle$ Fock sectors of the mesons with a complementary 3D confinement~\cite{Li:2015zda}. We solve this Hamiltonian in the leading two Fock components and fix the model parameters by fitting the known mass spectra of unflavored light mesons~\cite{Lan:2021wok}. The resulting light-front wavefunctions (LFWFs) obtained as eigenvectors of this Hamiltonian have been successfully employed to compute a wide class of pion observables, e.g., the electromagnetic form factors and associated charge radius, decay constant, quark and gluon PDFs, the differential cross sections for charmonium production by a pion beam, etc., with remarkable overall success. Here, we extend our investigations of the pion to compute its subleading-twist TMDs and predict, for the first time, the genuine twist-3 TMDs, $\tilde{e}$, which preserves the sum rule, and $\tilde{f}^\perp$ of the pion.

\section{Meson LFWFs in the BLFQ framework\label{Sec2}}
In light-front field theory, the solution of a bound state can be obtained by solving the light-front stationary Schr\"{o}dinger equation: $P^+P^-|\Psi\rangle=M^2|\Psi\rangle$, where $P^-$ and $P^+$ represent the light-front Hamiltonian and the longitudinal momentum of the system, respectively. At the fixed light-front time, $x^+\equiv x^0+x^3$, the meson state with mass squared eigenvalue $M^2$ can be expressed schematically in  terms of various quark, antiquark, and gluon Fock sectors, 
\begin{equation}
  |\Psi\rangle=\psi_{(q\bar{q})}|q\bar{q}\rangle+\psi_{(q\bar{q}g)}|q\bar{q}g\rangle+\psi_{(q\bar{q}q\bar{q})}|q\bar{q}q\bar{q}\rangle+\cdots,\label{psi0}
\end{equation}
where $\psi_{(\cdots)}$ is the LFWF associated with the Fock component $|\cdots\rangle$. For numerical calculations, we must truncate the infinite Fock sector expansion to a finite Fock space of Eq.~(\ref{psi0}), and here we retain the first and the second Fock components. This implies that the meson, at the model scale, is described by the quark-antiquark $\psi_{(q\bar{q})}$ and the quark-antiquark-gluon $\psi_{(q\bar{q}g)}$ LFWFs.


We adopt an effective light-front Hamiltonian: $P^-=P^-_{\text{QCD}}+P^-_C$~\cite{Lan:2021wok}, where $P^-_{\text{QCD}}$ and $P^-_C$ correspond to the light-front QCD Hamiltonian relevant to constituent $|q\bar{q}\rangle$ and $|q\bar{q}g\rangle$ Fock components of the mesons and a model Hamiltonian for the confining interaction, respectively. The light-front QCD Hamiltonian with one dynamical gluon in the light-front gauge $A^+=0$ is \cite{Lan:2021wok,Brodsky:1997de}
\begin{align}
    P_{\rm QCD}^-= &\int \mathrm{d}x^- \mathrm{d}^2 x^{\perp} \Big\{\frac{1}{2}\bar{\psi}\gamma^+\frac{m_{0}^2+(i\partial^\perp)^2}{i\partial^+}\psi\nonumber\\
    &+\frac{1}{2}A_a^i\left[m_g^2+(i\partial^\perp)^2\right] A^i_a +g\bar{\psi}\gamma_{\mu}T^aA_a^{\mu}\psi \nonumber\\
    &+ \frac{1}{2}g^2\bar{\psi}\gamma^+T^a\psi\frac{1}{(i\partial^+)^2}\bar{\psi}\gamma^+T^a\psi \Big\},
\end{align}
where $\psi$ and $A^\mu$ represent the quark and the gluon fields, respectively. $T^a$ is the half Gell-Mann matrix, $T^a=\lambda^a/2$. $g$ is the coupling constant. $m_0$ and $m_g$ are the bare quark mass and the model gluon mass, respectively. While the gluon mass is zero in QCD, we permit a phenomenologically motivated gluon mass to fit the mass spectra in our low-energy model~\cite{Lan:2021wok}. For the quark mass correction from a higher Fock sector, we introduce a mass counter term, $\delta m_q=m_0-m_q$, for the quark in the leading Fock sector, where $m_q$ is the renormalized quark mass. In the vertex interaction, we allow an independent quark mass $m_f$ referring to Ref.~\cite{Glazek:1992aq}.


The confinement in the leading Fock sector, which includes transverse and longitudinal confining potentials, reads as \cite{Lan:2021wok,Li:2015zda},
\begin{equation}
  \begin{split}\label{eqn:Hc}
  &P_{\rm C}^-P^+=\kappa^4\left\{x(1-x) \vec{r}_{\perp}^2-\frac{\partial_{x}[x(1-x)\partial_{x}]}{(m_q+m_{\bar{q}})^2}\right\},
  \end{split}
\end{equation}
where $\kappa$ is the strength of the confinement, and $\vec r_{\perp}=\sqrt{x(1-x)}(\vec r_{\perp q}-\vec r_{ \perp\bar{q}})$ represents the holographic variable \cite{Brodsky:2014yha}. Confinement in the $|q\bar{q}g\rangle$ sector is implemented through the cutoff of the BLFQ basis functions discussed below. However, since all mass eigenstates of the present model will have a valence quark-antiquark component, all our solutions will be confined due to the confining interaction in the quark-antiquark sector.

To compute the Hamiltonian matrix, we follow  BLFQ~\cite{Vary:2009gt} and consider the 2D harmonic-oscillator (HO) basis functions, $\Phi_{nm}(\vec{p}_\perp;b)$, with scale parameter $b$ to explain the transverse degrees of freedom~\cite{Li:2015zda}. The 2D-HO wave function carries the radial and the angular quantum numbers denoted by $n$ and $m$, respectively. We employ a plane-wave basis function to describe the longitudinal motion of the Fock particle. The longitudinal motion is confined to a 1D box of length $2L$ with antiperiodic (periodic) boundary conditions for fermion (boson). The longitudinal momentum is then defined as $p^+=\frac{2\pi}{L}k$, where $k$, an integer (half-integer) for boson (fermion), is the longitudinal quantum number. We neglect the zero mode for the boson. With the additional quantum number $\lambda$ for the light-front helicity, each Fock-particle basis state involves four quantum numbers $|\alpha_i\rangle=|k_i,n_i,m_i,\lambda_i\rangle$ and the many-body basis states are identified as the direct product of the Fock-particle basis states $|\alpha\rangle=\otimes_i|\alpha_i\rangle$. In the case of Fock sectors allowing multiple color-singlet states, we require an additional label to identify each color-singlet state. Note that this is only necessitated for the Fock sectors beyond the two that we retain here. In addition, our many-body basis states have well-defined total angular momentum projection $M_J=\sum_i (m_i+\lambda_i)$.

For numerical calculations, we truncate the infinite basis of each Fock sector by introducing two truncation parameters $K$ and $N_{\rm max}$ in longitudinal and transverse directions, respectively. The dimensionless variable $K=\sum_i k_i$ parameterizes the longitudinal momentum $P^+$. Thus, for a Fock particle $i$, the longitudinal momentum fraction $x$ is defined as $x_i=p^+/P^+=k_i/K$. The variable $K$ manifests as the `{\it resolution}' in the longitudinal direction, and hence a resolution onto the PDFs. The $N_{\rm max}$ truncation is set by $\sum_i(2n_i+|m_i|+1)\le N_{\rm max}$ and it enables factorization of the transverse center of mass motion \cite{Wiecki:2014ola,Zhao:2014xaa}. The $N_{\rm max}$ truncation implicitly acts as the infrared (IR) and ultraviolet (UV) cutoffs. In momentum space, the IR cutoff $\lambda_{\rm IR} \simeq b/\sqrt{N_{\rm max}}$ and the UV cutoff $\Lambda_{\rm UV} \simeq b\sqrt{N_{\rm max}}$ \cite{Zhao:2014xaa}. 

After diagonalizing the full Hamiltonian matrix in the BLFQ framework, we obtain the mass spectra $M^2$ and the corresponding LFWFs in momentum space,
\begin{align}
  &\Psi^{M_J}_{\mathcal{N},\{\lambda_i\}}({\{x_i,\vec{p}_{\perp i}\}})\nonumber\\
  &=\sum_{ \{n_i m_i\} }\psi^{M_J}_{\mathcal{N}}({\{\alpha}_i\})\prod_{i=1}^{\mathcal{N}}  \Phi_{n_i m_i}(\vec{p}_{\perp i},b)\,,
\label{eqn:wf}
\end{align}
where $\psi^{M_J}_{\mathcal{N}=2}(\{\alpha_i\})$ and $\psi^{M_J}_{\mathcal{N}=3}(\{\alpha_i\})$ are the components of the eigenvectors associated with the Fock sectors $|q\bar{q}\rangle$ and $|q\bar{q}g\rangle$, respectively.

With the truncations $\{N_{\text{max}},K\}=\{14,15\}$, the model parameters, summarized in Table \ref{para}, have been fixed to
generate the mass spectra of unflavored light mesons~\cite{Lan:2021wok}.
The LFWFs in this model provide a high-quality description of the electromagnetic form factors and associated charge radius, decay constant, PDFs for the pion, and pion-nucleus induced $J/\psi$
production cross sections~\cite{Lan:2021wok}.

\begin{table}[h]
  \caption{The model parameters for truncation $\{N_{\text{max}},K\}=\{14,15\}$~\cite{Lan:2021wok}. All are in units of GeV except $g$.}
  \vspace{0.15cm}
  \label{para}
  \centering
  \begin{tabular}{cccccc}
    \hline\hline
         $m_q$ & $m_g$ &$b$ &$\kappa$  &${m}_f$  &$g$ \\        
    \hline 
        0.39 & 0.60 &0.29 &0.65&5.69&1.92 \\       
    \hline\hline
  \end{tabular}
\end{table}

\section{Transverse-momentum-dependent parton distributions \label{Sec3}}
For spin-0 mesons, the quark TMDs are parameterized through the quark-quark correlation function defined as~\cite{Meissner:2008ay},
\begin{align}\label{coor}
      \Phi_q^{[\Gamma]}(x,k_\perp)=&\frac{1}{2}\int\frac{\mathrm{d}z^-\mathrm{d}^2\vec{z}_\perp}{2(2\pi)^3}e^{ik\cdot z}\nonumber\\
      &\times\langle P|\bar\psi(0)\Gamma\mathcal{W}(0,z)\psi(z)|P\rangle|_{z^+=0}
\end{align}
with $k^+=xP^+$. $|P\rangle$ is the light-front bound state of the target meson with mass $M$ and momenta $(P^+,\vec{P}_\perp)$, where the transverse momentum $\vec{P}_\perp=\vec{0}$ \cite{Collins:1992kk}. The Dirac matrix $\Gamma$ determines the Lorentz structure of the correlator $\Phi^{[\Gamma]}$ and its `twist' $\tau$ \cite{Jaffe:1991kp}. The Wilson line $\mathcal{W}$ preserves the gauge invariance of the bilocal quark field operators in the correlation function \cite{Bacchetta:2020vty}. 

For the pion, there are two leading twist (twist-2) TMDs namely the unpolarized quark TMD, $f^q_{1}(x,k_\perp)$, and the polarized quark TMD, $h_{1}^{\perp q}(x,k_\perp)$, also known as the Boer-Mulders function, which are defined through the parameterizations of the quark-quark correlator with $\Gamma\equiv\gamma^{+},\,\sigma^{j+} \gamma_{5}$,
\begin{align}
  \Phi_q^{[\gamma^{+}]}(x,k_\perp) &=f^q_{1}(x,k_\perp), \label{tw-2TMDs1}\\
  \Phi_q^{[i \sigma^{j+} \gamma_{5}]}(x,k_\perp) &=-\frac{ \varepsilon_{T}^{i j} k_{\perp}^i}{M} {h_{1}^{\perp q}}(x,k_\perp)\label{tw-2TMDs2},
\end{align}
where $k_\perp$ is the transverse momentum of a struck quark, $\epsilon_T^{11}=\epsilon_T^{22}=0$, and $\epsilon_T^{12}=-\epsilon^{21}=1$. 
Meanwhile, the subleading twist (twist-3) quark TMDs are defined as follows 
\begin{align}
    \Phi^{[1]}_q (x,k_\perp)&=\frac{M}{P^{+}}e^q(x,k_\perp),     \label{tw-3TMDs-e}\\
    \Phi_q^{[\gamma^{j}]}(x,k_\perp) &=\frac{M}{P^{+}} \frac{k_{\perp}^{j}}{M}f^{\perp q}(x,k_\perp), \label{tw-3TMDs-fperp}\\
    \Phi_q^{[\gamma^{j} \gamma_{5}]}(x,k_\perp) &=-\frac{M}{P^{+}}\frac{\varepsilon_{T}^{i j} k_{\perp}^{i}}{M}{g^{\perp q}}(x,k_\perp), \\
    \Phi_q^{[i \sigma^{i j} \gamma_{5}]}(x,k_\perp) &=-\frac{M}{P^{+}} \varepsilon_{T}^{i j} {h}^q(x,k_\perp)\label{tw-3TMDs-h}. 
\end{align}
The TMDs $h_{1}^{\perp q}(x,k_\perp)$, $g^{\perp q}(x,k_\perp)$ and $h^q(x,k_\perp)$ are T-odd TMDs in nature. The rest are T-even TMDs \cite{Meissner:2008ay}. Note that the twist-3 TMD correlation functions have a suppression with the factor $M/P^+$. 
In this present work, we do not consider the effect of the Wilson line $\mathcal{W}$ and set it to be a unit matrix. With this approximation: $\mathcal{W}\approx\mathbbm{1}$, only the T-even TMDs survive~\cite{Lorce:2016ugb}.

In the light-front field theory~\cite{Kogut:1969xa}, the quark field is decomposed into a `good' $\psi_+$, and a `bad' $\psi_-$ component ($\psi_{\pm}=\frac{1}{4}\gamma^{\mp}\gamma^{\pm}\psi$)\footnote{The projection operators used here are different from the definition in Ref.~\cite{Kogut:1969xa}. We follow the convention, $x^\pm\equiv x^0\pm x^3$.}. The bad component $\psi_-$ can be expressed in terms of the good component $\psi_+$ and the gauge fields as,
\begin{equation}
    \psi_-(z)=\frac{\gamma^+}{2i\partial^+}\left[i(\partial_j-igA_{\perp j}(z))\gamma_j+m_q\right]\psi_+(z),
\end{equation}
where $m_q$ is the quark mass, and $j=1,2$. The above constraint equation comes from the EOM in the light-cone gauge $A^+=0$. Note that  in Eqs.~(\ref{tw-2TMDs1}) and (\ref{tw-2TMDs2}), the Dirac matrices in the leading-twist operators project the correlator into $\bar{\psi}_+\psi_+$, while in Eqs.~(\ref{tw-3TMDs-e})-(\ref{tw-3TMDs-h}), the Dirac matrices in the subleading-twist operators project the correlator into $\bar{\psi}_+\psi_-+\bar{\psi}_-\psi_+$. Therefore, the twist-3 TMDs can not be expressed as the probabilities or the differences of probabilities. They are related to quark-quark matrix elements and quark-quark-gluon matrix elements. 


The higher-twist operators can be decomposed into the twist-2 and other operators using the EOM of quark fields \cite{Efremov:2002qh}. Consequently, the twist-2 (Eqs.~(\ref{tw-2TMDs1}) and (\ref{tw-2TMDs2})) and the twist-3 (Eqs.~(\ref{tw-3TMDs-e})-(\ref{tw-3TMDs-h})) TMDs are related via the following relations~\cite{Lorce:2016ugb} 
\begin{align}
    xe^q(x,k_\perp)&=x\tilde{e}^q(x,k_\perp)+\frac{m_q}{M}f^q_1(x,k_\perp),\label{EOMeom1}\\
    xf^{\perp q}(x,k_\perp)&=x\tilde{f}^{\perp q}(x,k_\perp)+f^q_1(x,k_\perp),\label{EOMeom2}
\end{align}
where the tilde terms or the so-called genuine twist-3 TMDs are defined as follows (see appendix for detailed derivation)
\begin{align}
      &\tilde{e}^q(x,k_\perp)=\frac{i}{Mx}\frac{g}{2}\int\frac{\mathrm{d}z^-\mathrm{d}^2\vec{z}_\perp}{2(2\pi)^3}e^{ik\cdot z}\nonumber\\
      &\times\langle P|\bar{\psi}_+(0)\sigma^{j+}[A_\perp^j(z)-A^j_\perp(0)]\psi_+(z)|P\rangle|_{z^+=0},\\
      &2xk^j_\perp\tilde{f}^{\perp q}(x,k_\perp)=g\int\frac{\mathrm{d}z^-\mathrm{d}^2\vec{z}_\perp}{2(2\pi)^3}e^{ik\cdot z}\nonumber\\
      &\times\langle P|\bar{\psi}_+(0)\gamma^+[A_\perp^j(0)+A_\perp^j(z)]\psi_+(z)|P\rangle |_{z^+=0}.
\end{align}

Sum rules are particularly important for testing the model consistency. 
The twist-2 and the twist-3 integrated TMDs, $j(x)=\int\mathrm{d}^2{k}_\perp j(x,k_\perp)$, where $j(x,k_\perp)$ denotes the TMDs, must satisfy the following sum rules \cite{Lorce:2016ugb}
\begin{align}
  \int\mathrm{d}x f^q_1(x)&=N_q,\label{quarkNum}\\
  \sum_q\int\mathrm{d}x xf_1^q(x)&=1-\int\mathrm{d}x xf_1^g(x)\label{momSumRul},\\
  \sum_q\int\mathrm{d}x e^q(x)&=\frac{\sigma_\pi}{m_q},\label{scalarFF}\\
  \int\mathrm{d}x xe^q(x)&=\frac{m_q}{M}N_q.\label{firstmoment}
\end{align}
Equation~\eqref{quarkNum} represents the number sum rule, where $N_q$ is the total number of valence quarks of a particular species \cite{Schwartz:2014sze}. 
Equation~\eqref{momSumRul} corresponds to the momentum sum rule, which states that the constituent quark, antiquark, and gluon together carry the entire light front momentum of the meson. 
Note that due to the explicit $k_\perp^j$ factor in Eq.~(\ref{tw-3TMDs-fperp}), there does not exist any PDF nor sum rules corresponding to the TMD $f^{\perp q}(x,k_\perp)$.
Since the focus of the present work is on the quark distributions, we omit the superscript $q$ from the TMDs below for simplicity.


The zeroth moment of $e(x)$, Eq.~(\ref{scalarFF}), involves the singular term at $x=0$ and is related to the scalar form factor of the pion at zero momentum transfer, $\sigma_{\pi}$ \cite{Jaffe:1991ra,Efremov:2002qh}. 
The second sum rule of $e(x)$, Eq.~(\ref{firstmoment}), can be cast into the first-moment sum rule of $\tilde{e}(x)$ using the relation of TMDs at different twists in Eq.~(\ref{EOMeom1}) and the number sum rule given in Eq.~(\ref{quarkNum}),
\begin{equation}
  \int\mathrm{d}x\, x\,\tilde{e}(x)=0.\label{SumRul_xetilde}
\end{equation}
We will discuss this sum rule within our model in the following section.


\subsection*{Overlap representations of TMDs\label{Sec4}}
Here, we provide an overview of the LFWF overlap representations for the leading-twist and the subleading-twist TMDs of the pion. 
In the Fock space, the meson state in Eq.~\eqref{coor} is expressed by considering all Fock components as    
\begin{align}
  |P\rangle&=\sum_n\sum_{\lambda_{1},\lambda_{2},\cdots,\lambda_{n}} \int \prod^n_{1} \frac{\left[\mathrm{d}x_i\mathrm{d}^2\vec{k}_{\perp i}\right]}{2(2\pi)^3\sqrt{x_i}} \nonumber\\
  &\times2(2\pi)^3\delta\left(1-\sum_1^nx_i\right)\delta^2\left(\vec{P}_\perp-\sum_1^n\vec{k}_{\perp i}\right)\nonumber\\
  &\times\Psi_{n,\{\lambda_i\}}\left(\left\{x_{i}, k_{\perp i}\right\}\right)\left|\{x_i P^+,\vec k_{\perp i} +x_i \vec P_{\perp},\lambda_i\}\right\rangle ,
    \label{BoundState0}
\end{align}
with the following orthonormal relations,
\begin{equation}
  \langle P^\prime|P\rangle=\sum_{n,n^\prime}2P^+(2\pi)^3\delta^3(P-P^\prime)\delta_{nn^\prime}P_n,
\end{equation}
where $\delta^3(P-P^\prime)=\delta( P^+- P^{\prime+})\delta^2(\vec P_\perp-\vec P_\perp^\prime)$. The summation is over all the Fock components and $P_n$ defines the probability of finding the $n$-parton state in the meson. In this work, we have only two LFWFs, $\psi_2$ for $|q\bar{q}\rangle$ and $\psi_3$ for $|q\bar{q}g\rangle$. With the model parameters summarized in Table \ref{para}, we find that the probabilities of the $|q\bar{q}\rangle$ and $|q\bar{q}g\rangle$ components are $49.2\%$ and $50.8\%$, respectively.

Using the expression of the meson state, Eq.~(\ref{BoundState0}), and the quantized quark and gluon fields in the Lepage-Brodsky prescription~\cite{Brodsky:1997de,Lepage:1980fj}, we express the TMDs $f_1(x,k_\perp)$, $\tilde{e}(x,k_\perp)$, and $\tilde{f}^\perp(x,k_\perp)$ in terms of the overlap of the LFWFs as 
\begin{align}
  f_1(x,k_\perp&)=\sum_{\lambda_1,\lambda_2} \mathrm{d}[12]\Psi^*_{2,\lambda_1\lambda_2}\Psi_{2,\lambda_1\lambda_2}\delta^3(\tilde{p}^\prime_1-\tilde{k})\nonumber\\
  +\sum_{\lambda_1,\lambda_2,\lambda_g}&\int \mathrm{d}[123] \Psi^*_{3,\lambda_1\lambda_2\lambda_g}\Psi_{3,\lambda_1\lambda_2\lambda_g}\delta^3(\tilde{p}_1-\tilde{k}),\label{f1_overlaps}\\
  \tilde{e}(x,k_\perp&)=-\frac{gC_F\sqrt{2}}{Mx}\sum_{\lambda_2,\lambda_g}\int\mathrm{d}[12]\mathrm{d}[123]\frac{2(2\pi)^3}{\sqrt{x_g}}\nonumber\\
  &\times(\Psi^*_{2,-\lambda_2}\Psi_{3,+\lambda_2\lambda_g}-\Psi_{2,+\lambda_2}\Psi^*_{3,-\lambda_2\lambda_g})\notag\\
  &\times[\delta^3(\tilde{p}^\prime_1-\tilde{k})-\delta^3(\tilde{p}_1-\tilde{k})]\delta^3(\tilde{p}^\prime_2-\tilde{p}_2),\\
  \tilde{f}^\perp(x,k_\perp&)=\frac{gC_F\sqrt{2}}{2xk^1_\perp}\sum_{\lambda_1,\lambda_2,\lambda_g}\int\mathrm{d}[12]\mathrm{d}[123]\frac{2(2\pi)^3}{\sqrt{x_g}}(-)^{\frac{\lambda_g+1}{2}}\notag\\
  &\times[\Psi^*_{2,\lambda_1\lambda_2}\Psi_{3,\lambda_1\lambda_2\lambda_g}\delta^3(\tilde{p}_1-\tilde{k})\notag\\
  &+\Psi_{2,\lambda_1\lambda_2}\Psi^*_{3,\lambda_1\lambda_2,\lambda_g}\delta^3(\tilde{p}^\prime_1-\tilde{k})]\delta^3(\tilde{p}_2^\prime-\tilde{p}_2)\label{fperp_overlaps}.
\end{align}
Here we define the conventions
\begin{align}
  \mathrm{d}[12]\equiv \frac{\prod_{1}^2\mathrm{d}x^\prime_{i}\mathrm{d}^2\vec{p}^{\;\prime}_{\perp i}}{[2(2\pi)^3]^{2}}2(2\pi)^3\delta^3(\tilde{P}-\sum_{1}^2\tilde{p}^\prime_i),\\
  \mathrm{d}[123]\equiv \frac{\prod_{1}^3\mathrm{d}x_{i}\mathrm{d}^2\vec{p}_{\perp i}}{[2(2\pi)^3]^{3}}2(2\pi)^3\delta^3(\tilde{P}-\sum_{1}^3\tilde{p}_i),
\end{align}
where $\delta^3(\tilde{P}-\sum_1^n\tilde{p}_i)=\delta(1-\sum_1^nx_i)\delta^2(\vec{P}_\perp-\sum_1^n\vec{p}_{\perp i})$, and we omit the arguments $(\tilde{p}^\prime_1,\tilde{p}^\prime_2)$ from the quark-antiquark LFWFs and $(\tilde{p}_1,\tilde{p}_2,\tilde{p}_g)$ from the quark-antiquark-gluon LFWFs, where $\tilde{k}\equiv(x,\vec{k}_\perp)$. The color factor, $C_F={2}/{\sqrt{3}}$, comes from the quark-quark-gluon matrix element $ \langle q\bar{q}|\bar{\psi}_{b}A^{\mu,a}T^a_{bc}\psi_c|q\bar{q}g\rangle$ and its hermitian conjugate. The unpolarized TMD $f_1(x,k_\perp)$ involves the individual overlaps of LFWFs of the $|q\bar{q}\rangle$ and the $|q\bar{q}g\rangle$ Fock sectors. However, the genuine twist-3 TMDs $\tilde{e}(x,k_\perp)$ and $\tilde{f}^\perp(x,k_\perp)$ involve the mixed overlaps of LFWFs between the $|q\bar{q}\rangle$ and the $|q\bar{q}g\rangle$ Fock sectors. 
This is interpreted as an interference between scattering from a coherent quark-gluon pair and from a single quark \cite{Jaffe:1991kp,Jaffe:1991ra,Efremov:2002qh}.

\section{Numerical results \label{Sec5}}

With the model parameters for the basis truncation $\{N_{\text{max}},K\}=\{14,15\}$ summarized in Table \ref{para}, we solve the light-front stationary Schr\"{o}dinger equation to obtain the two Fock-state LFWFs of the pion with the mass $M=0.139~\mathrm{GeV}$.
The LFWFs $\Psi_{n,\{\lambda_i\}}(\{x_i,\vec{k}_{\perp i}\})$ are boost-invariant and depend only on the longitudinal fraction $x_{i}$ and the relative transverse momentum of the constituent parton of the meson $\vec{k}_{\perp i}$. The physical transverse momentum of the parton is given by $\vec{ p}_{\perp i} = \vec{k}_{\perp i}+x_{i} \vec{ P}_{\perp}$ and the longitudinal momentum is $p_{i}^+ = k_{i}^+ = x_{i}P^+$. 
We convert the nonperturbative solutions in the single particle coordinates, Eq.~(\ref{eqn:wf}), to those in the relative coordinates, $\Psi_{n,\{\lambda_i\}}(\{x_i,\vec{k}_{\perp i}\})$, by factorizing out the center of mass motion~\cite{Wiecki:2014ola,Xu:2021wwj,Hu:2020arv}.
We then employ the resulting LFWFs with the expressions in Eqs. (\ref{f1_overlaps}-\ref{fperp_overlaps}) to compute the quark TMDs and PDFs of the pion.

\subsection{Twist-2 and twist-3 TMDs} 
The TMDs computed in BLFQ have oscillations in the transverse momentum direction due to the oscillatory nature of the HO basis functions applied in the transverse plane. The transverse truncation $N_{\rm max}$, together with the HO scale $b$ determines the transverse basis states. The average over the BLFQ results at different $N_{\rm max}$ decreases these finite basis artifacts. Following the averaging procedure adopted in Refs.~\cite{Hu:2020arv,Nair:2022evk}, we take an average of the BLFQ results for three different $N_{\mathrm{max}}$ at fixed $K$. 
We first take the average between the BLFQ results at $N_{\mathrm{max}} = n$ and $N_{\mathrm{max}} = n + 2$. We consider another average of the results obtained at $N_{\mathrm{max}} = n+2$ and $N_{\mathrm{max}} = n + 4$. The final result is then achieved by averaging the previous two averages. Thus, this two-step averaging method involves the BLFQ results at $N_{\mathrm{max}} = \{n,\, n + 2,\,n+4\}$, and at fixed $K$.

\begin{figure*}[h]
  \centering
      \includegraphics[width=0.3\textwidth]{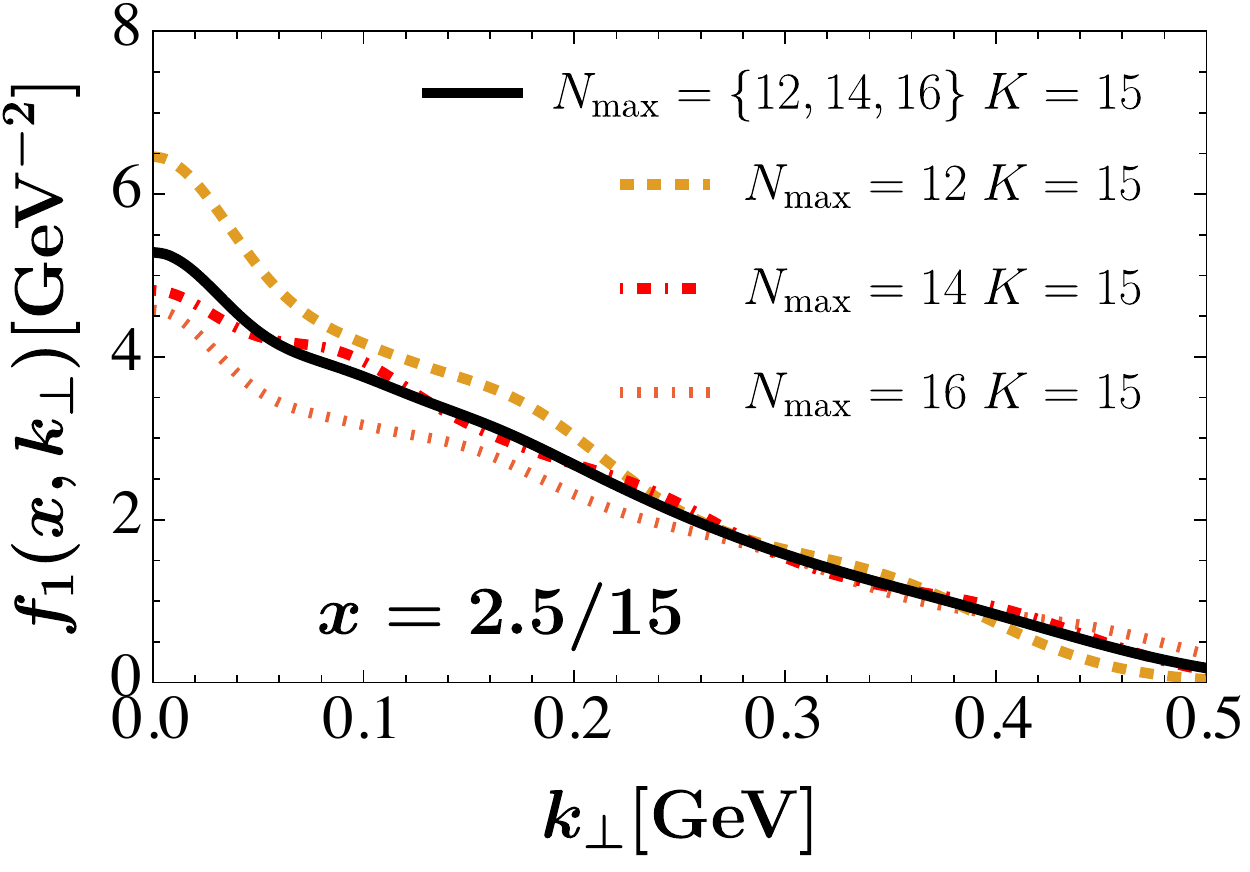}  
      \includegraphics[width=0.31\textwidth]{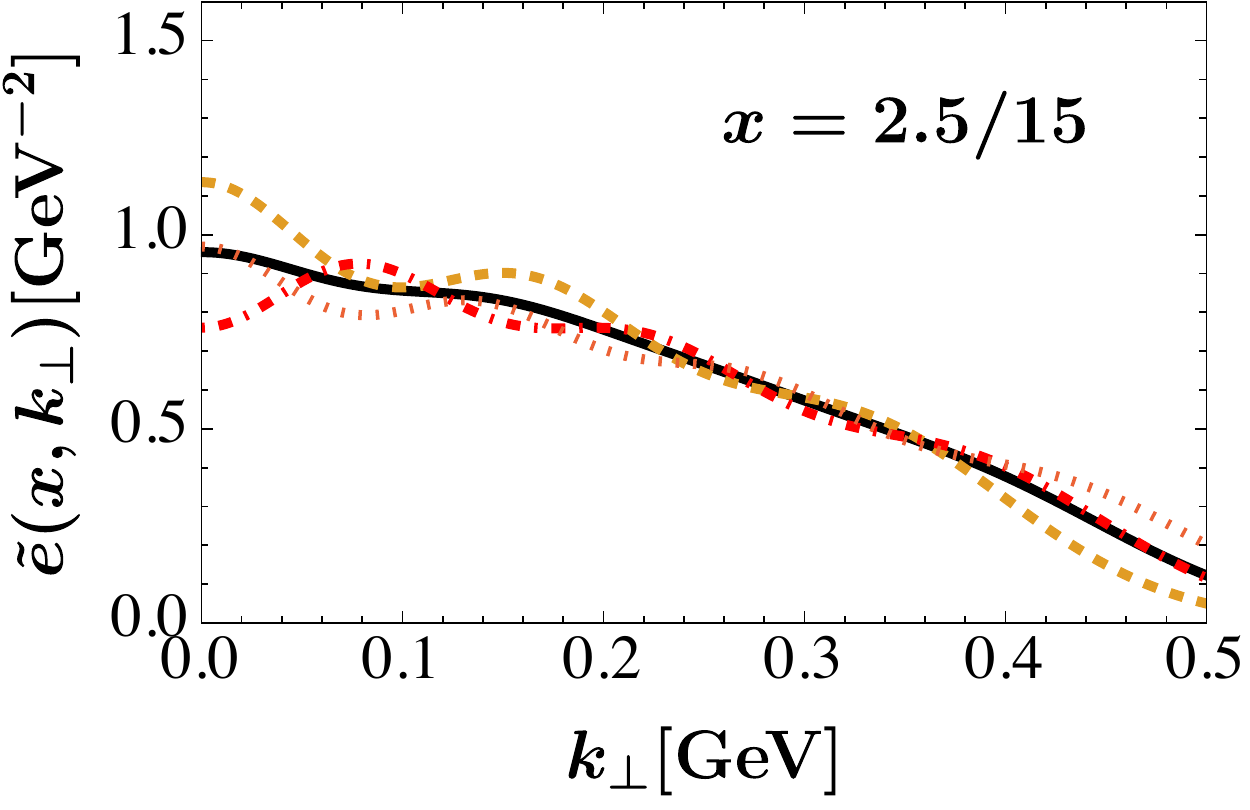}  
      \includegraphics[width=0.31\textwidth]{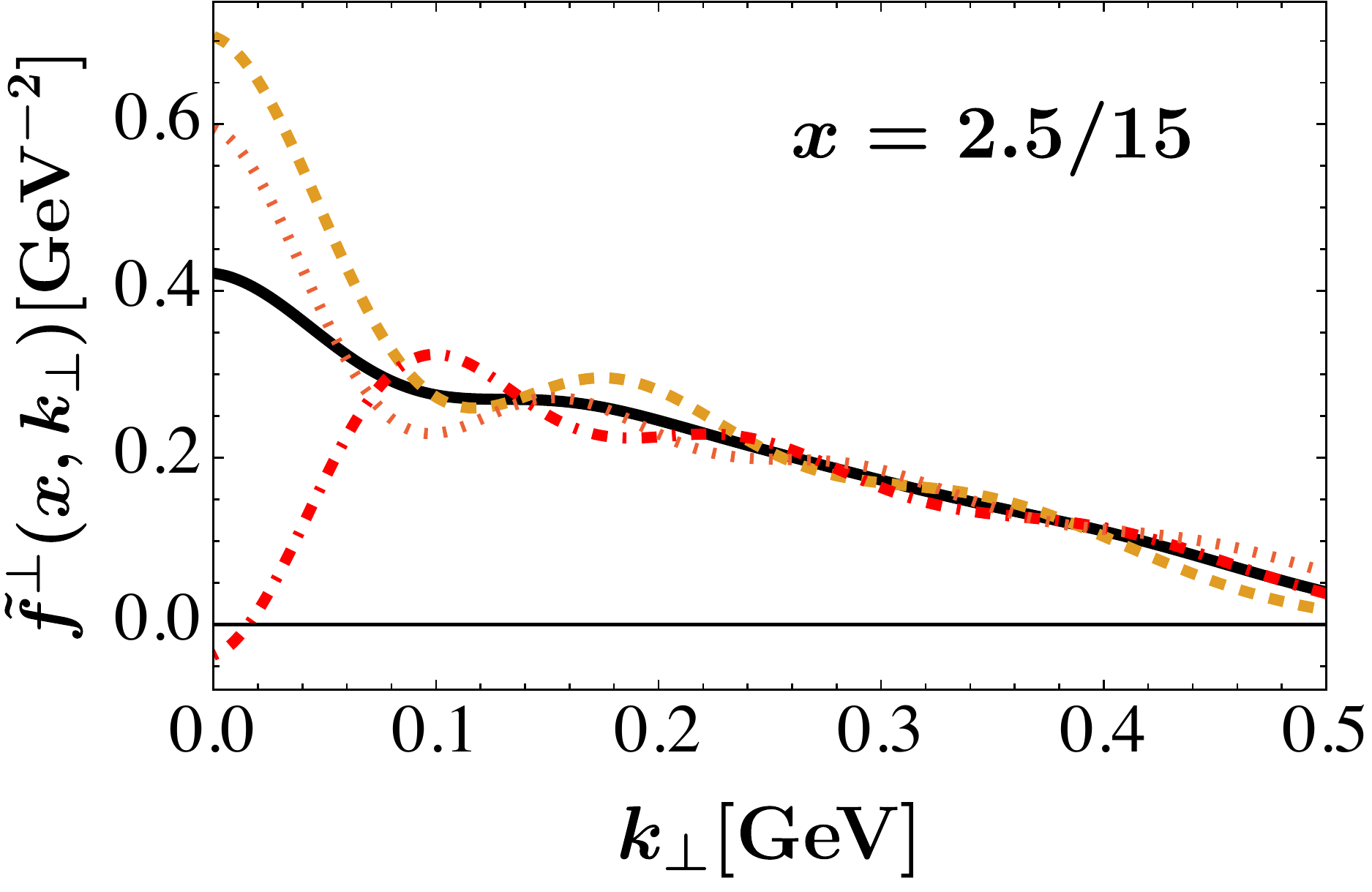} \\
      \includegraphics[width=0.3\textwidth]{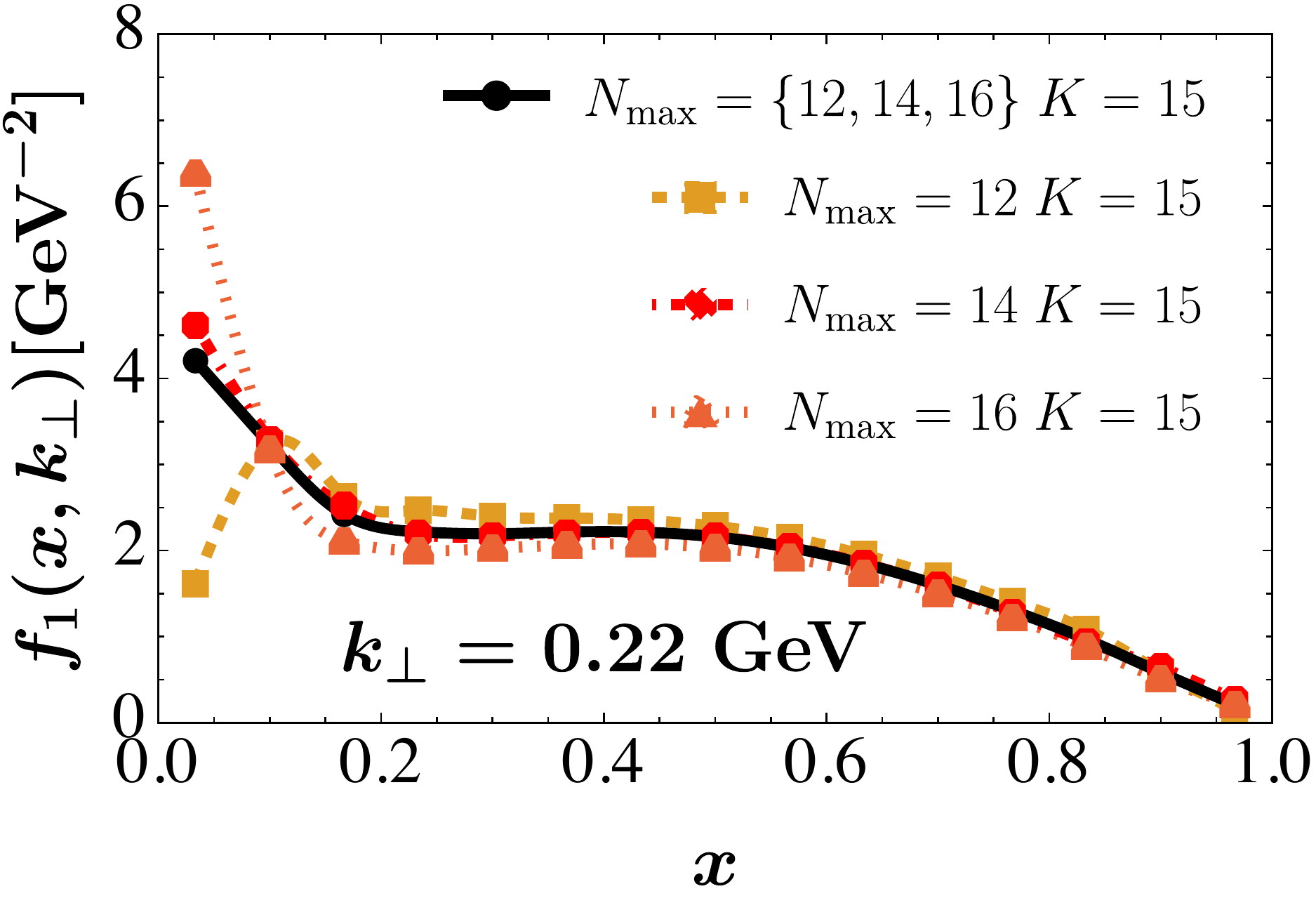}  
      \includegraphics[width=0.3\textwidth]{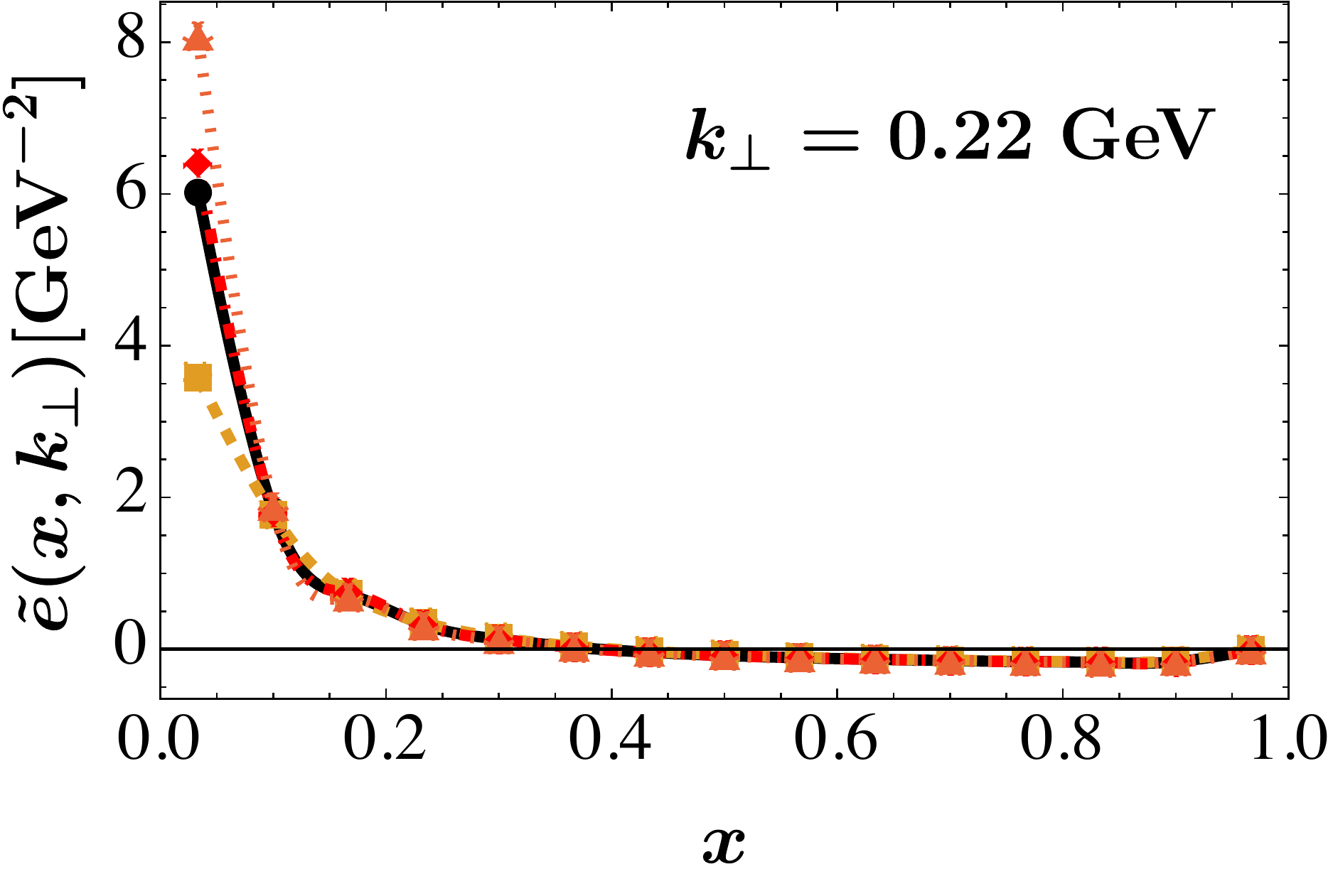}  
      \includegraphics[width=0.31\textwidth]{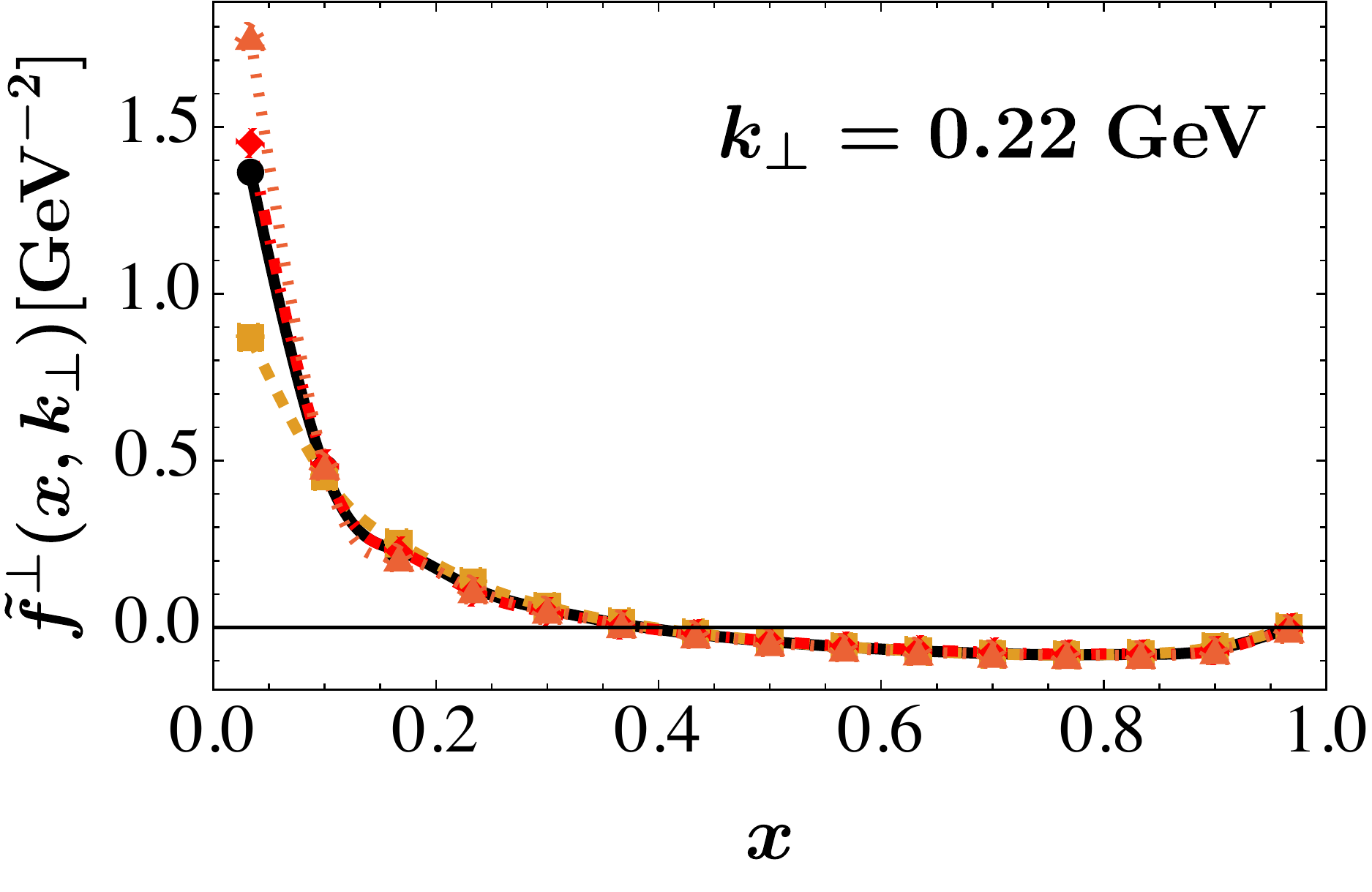}   
      \caption{The BLFQ results for the twist-2 TMD, $f_1(x,k_\perp)$, and the genuine twist-3 TMDs, $\tilde{e}(x,k_\perp)$ and $\tilde{f}^\perp(x,k_\perp)$ at different truncation parameters. Upper panel: the TMDs are functions of $k_\perp$ for fixed $x=2.5/15$. Lower panel: the TMDs are functions of $x$ for fixed $k_\perp=0.22$ GeV. The solid black lines represent the TMDs after implementing the averaging scheme with truncation parameters $N_{\text{max}}=\{12,14,16\}$, $K=15$. 
      }\label{TMDs3D2D}
\end{figure*}

Figure~\ref{TMDs3D2D} shows the pion's TMDs with different $N_{\mathrm{max}}$ and $K=15$. The upper panel of Fig.~\ref{TMDs3D2D} presents the TMDs as functions of $k_\perp$ at fixed longitudinal momentum fraction $x=2.5/15$, whereas the lower panel of Fig.~\ref{TMDs3D2D} shows the TMDs as functions of $x$ at fixed $k_\perp=0.22$ GeV.  
We observe that the BLFQ results at different $N_{\mathrm{max}}$ have oscillations. The magnitude of the oscillation is larger when $k_\perp$ and $x$ are small. The phases of the oscillations alter with changing $N_{\mathrm{max}}$.
Through the above method, we average the BLFQ results of different $N_{\mathrm{max}}$, which reduces the oscillation of our results. This can be seen in Fig.~\ref{TMDs3D2D} (solid black lines). Meanwhile, the amplitude of the oscillations is small in the large $k_\perp$ and $x>0.1$ region reflecting better stability of our results in that domain.

\begin{figure*}[h]
  \centering
      \includegraphics[width=0.32\textwidth]{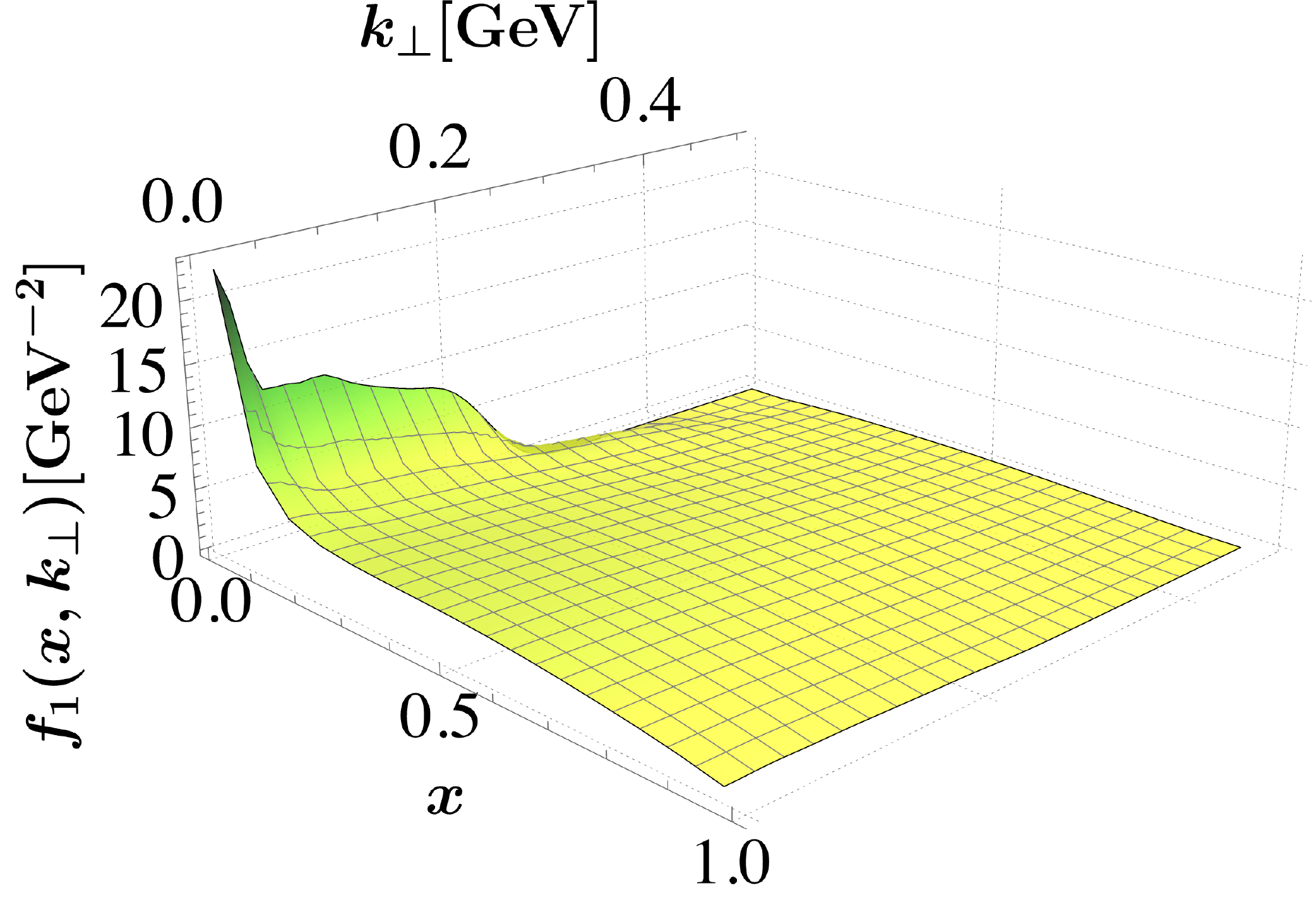}  
      \includegraphics[width=0.32\textwidth]{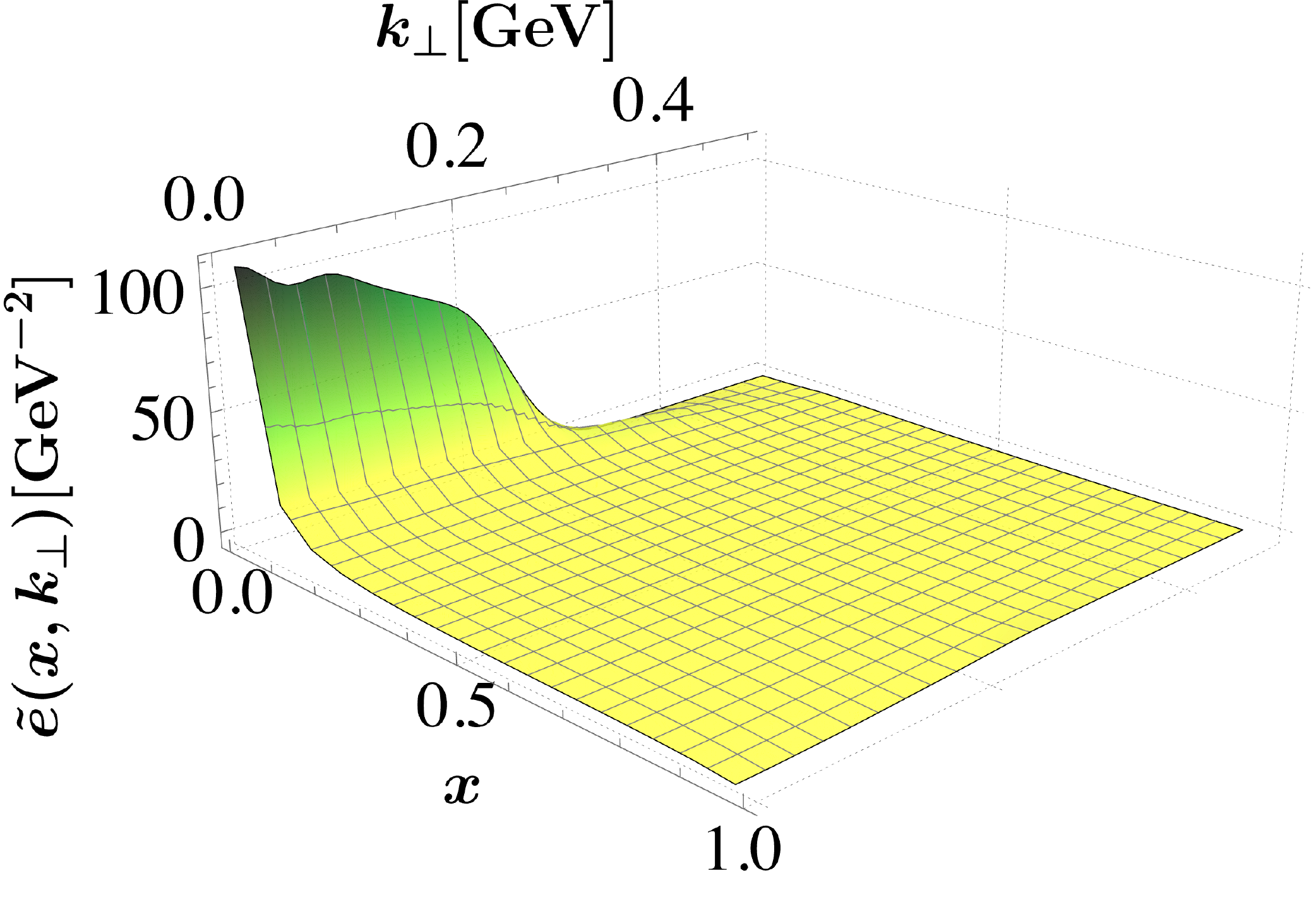}    
      \includegraphics[width=0.32\textwidth]{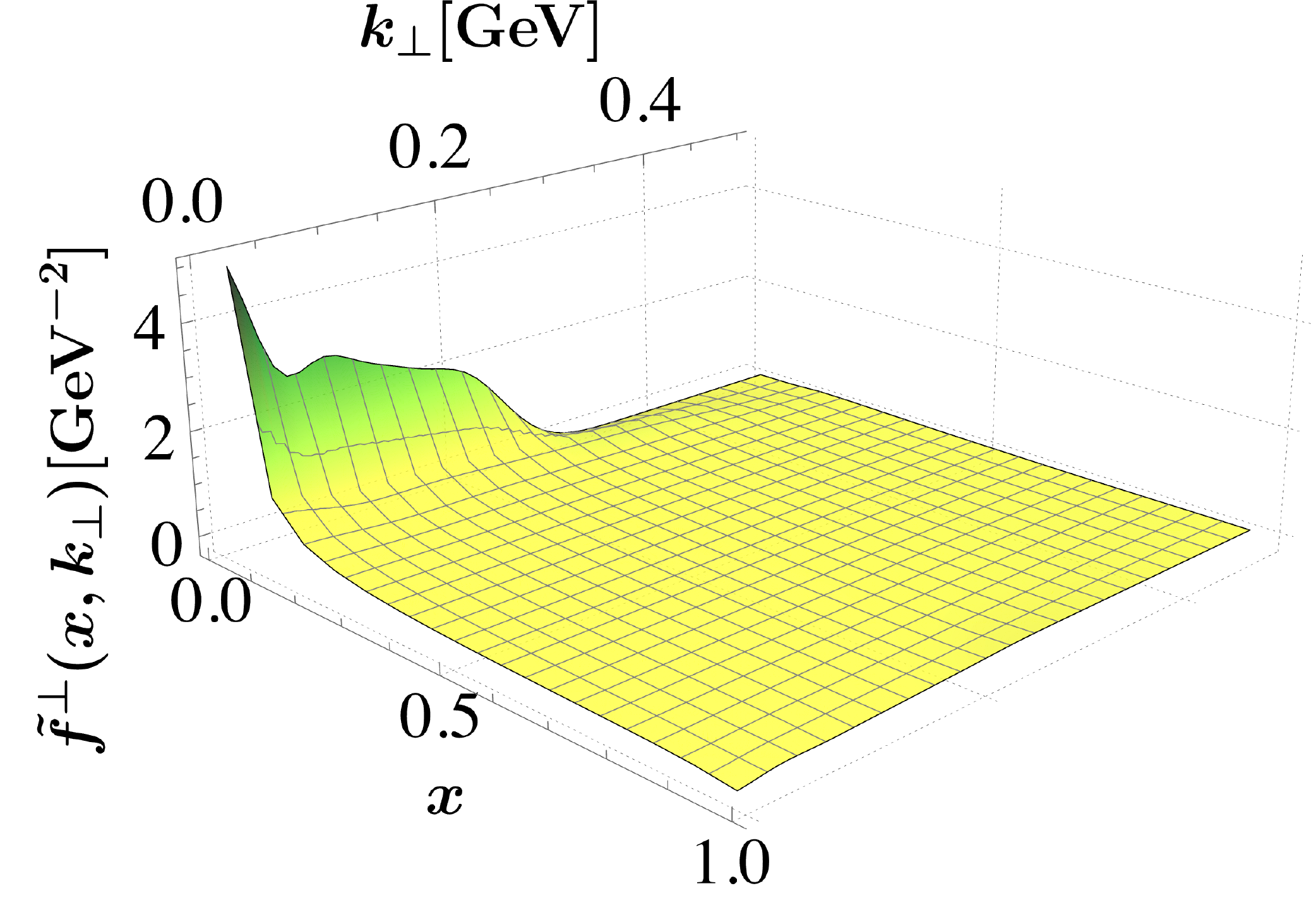}   
      \caption{3D plots of the BLFQ results for the twist-2 TMD, $f_1(x,k_\perp)$, and the genuine twist-3 TMDs, $\tilde{e}(x,k_\perp)$ and $\tilde{f}^\perp(x,k_\perp)$. Results are all obtained by averaging over the BLFQ results at $N_{\text{max}}=\{12,14,16\}$, $K=15$.
      }\label{TMDs3D}
\end{figure*}

The 3D structure of our BLFQ results for the TMDs $f_1(x,k_\perp)$, $\tilde{e}(x,k_\perp)$, and $\tilde{f}^\perp(x,k_\perp)$, after implementing the average, is shown in Fig.~\ref{TMDs3D}. We find that the TMDs still have small oscillations in the $k_\perp$ direction after averaging. The TMDs drop to zero abruptly in the transverse plane after the UV cutoff ($\sim b\sqrt{N_{\text{max}}}$). 
Note that the UV cutoff increases with increasing $N_{\text{max}}$ and thus, the fall-off of the TMDs extends to the relevant higher-momentum region for larger $N_{\text{max}}$. This behavior of TMDs within our BLFQ framework has been observed for the physical spin-$\frac{1}{2}$ particle in QED, i.e., electrons reported in Ref.~\cite{Hu:2020arv}.

\begin{figure}[h] 
  \centering
    \includegraphics[width=0.32\textwidth]{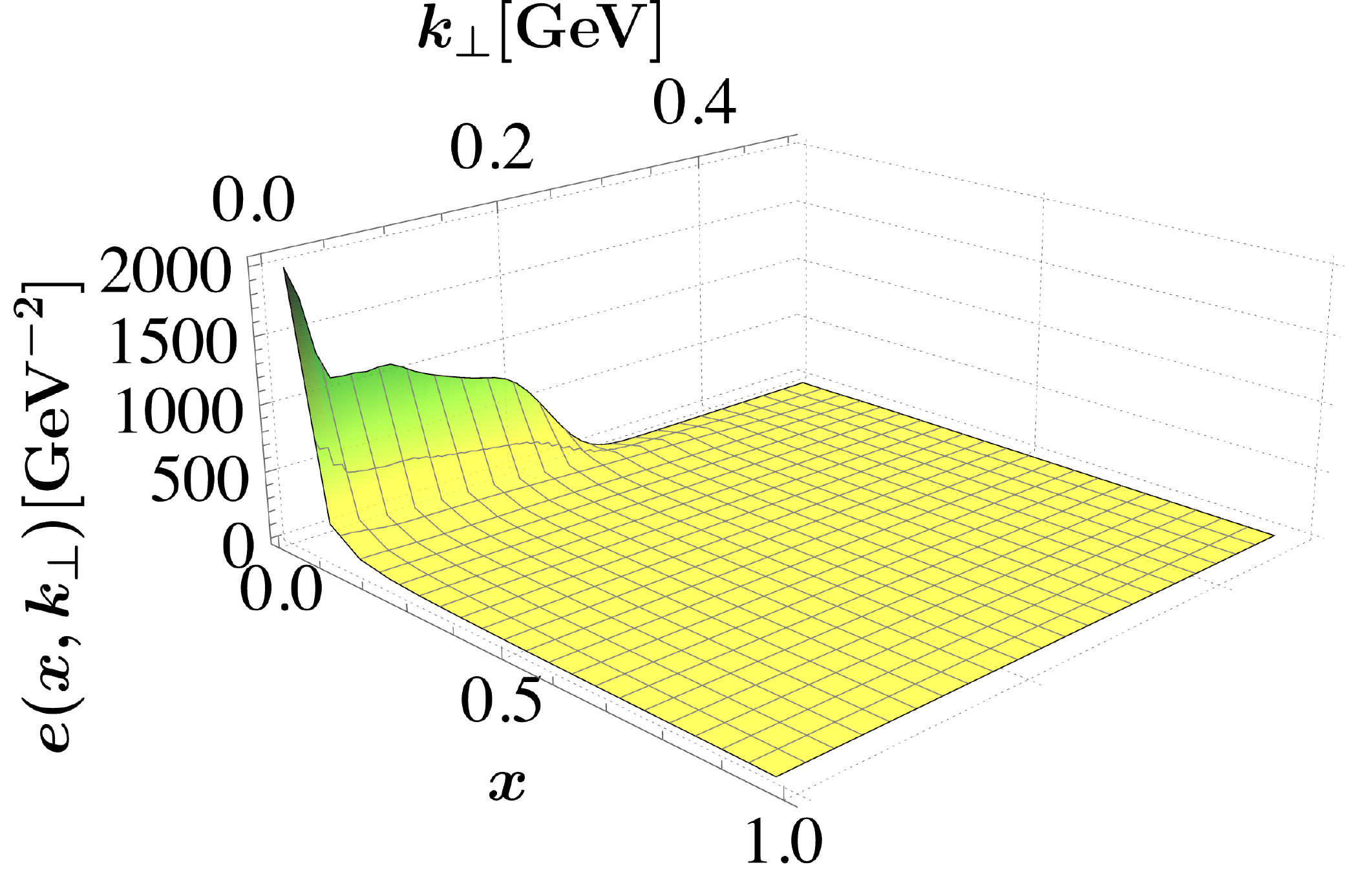} 
    \includegraphics[width=0.32\textwidth]{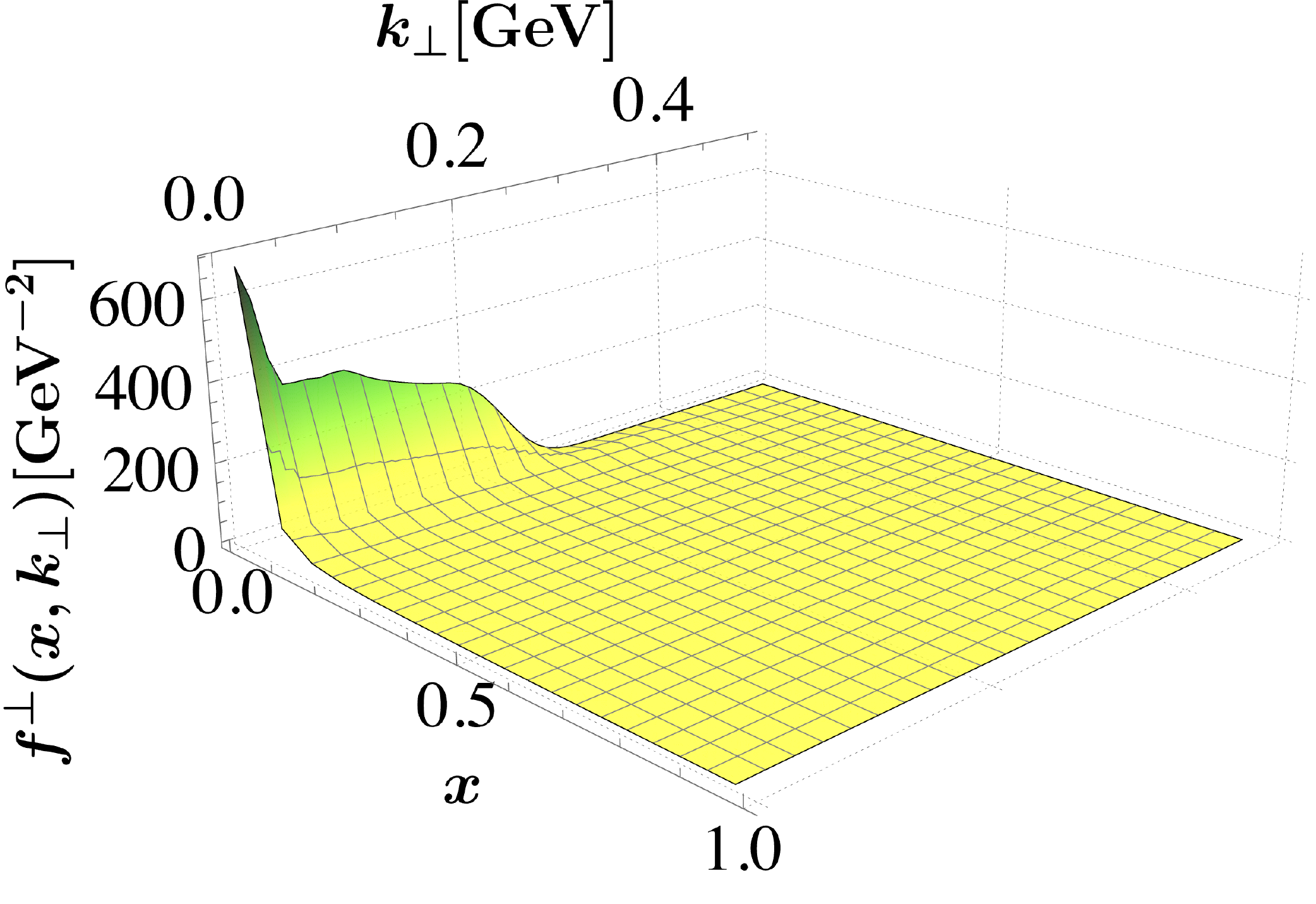}  
    \caption{3D plots of the BLFQ results for the twist-3 TMDs, $e(x,k_\perp)$ and $f^\perp(x,k_\perp)$. Results are all obtained by averaging over the BLFQ results at $N_{\text{max}}=\{12,14,16\}$, $K=15$.
    }
    \label{tw-3TMDs}
\end{figure}

Figure \ref{tw-3TMDs} presents the total twist-3 TMDs, $e(x,k_\perp)$ and $f^\perp(x,k_\perp)$, using the EOM relations given in Eqs. (\ref{EOMeom1}) and (\ref{EOMeom2}). We find that the qualitative behavior of these TMDs is similar to each other and their magnitudes are larger than that of the twist-2 TMD $f_1(x,k_\perp)$. The behavior of these TMDs mainly follows from the function $f_1(x,k_\perp)/x$, which is singular at $x=0$. However, the contributions of twist-3 TMDs in the cross section are suppressed by a factor $M/P^+$ \cite{Bacchetta:2019qkv} and thus, the higher twist effects are weakened in high-energy collisions. It can also be noticed that the twist-3 TMDs fall quickly in the transverse direction.

\subsection{$k_\perp$ moments of the TMDs}
\begin{figure}[h]
\centering
  \includegraphics[width=0.32\textwidth]{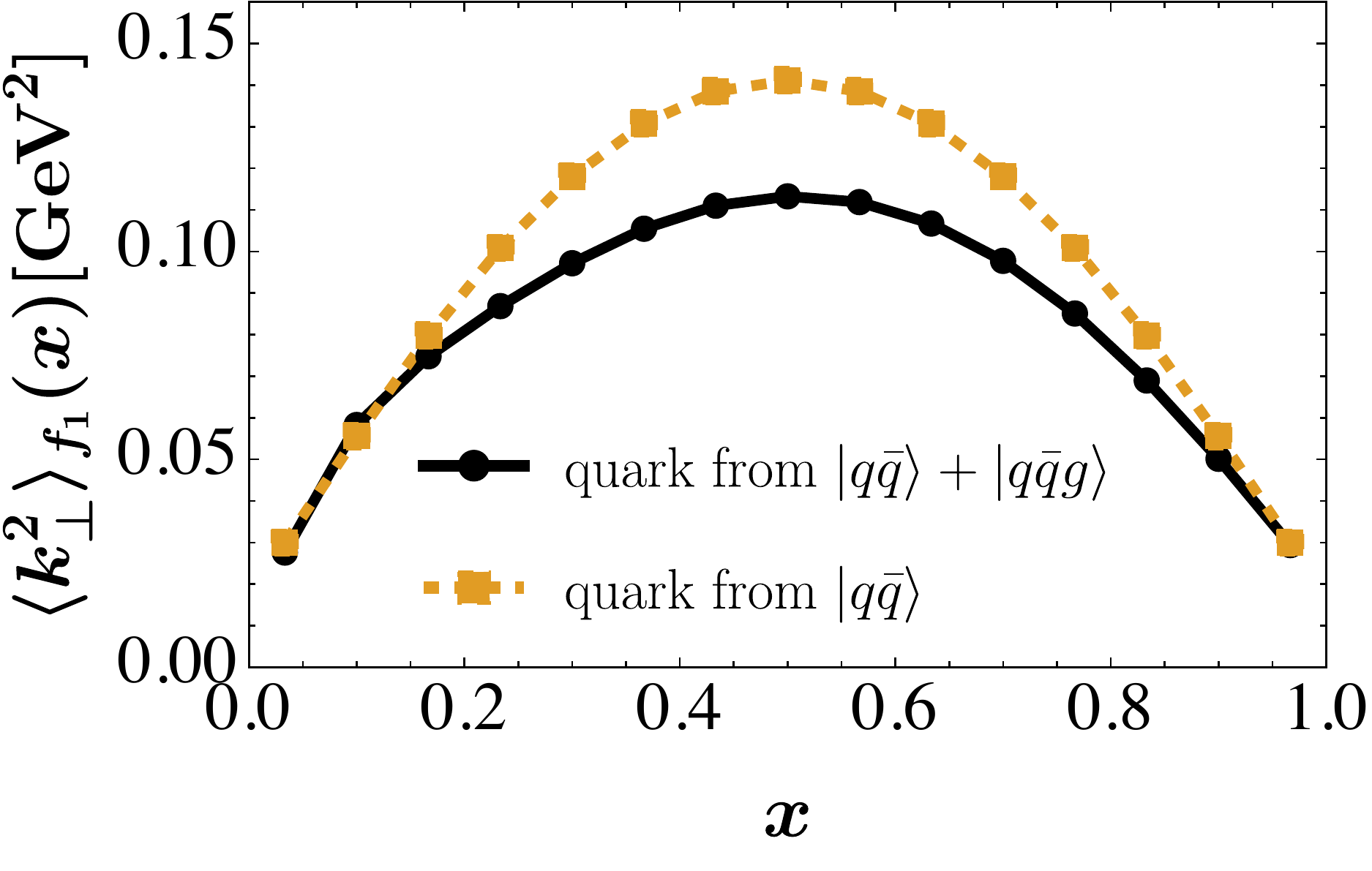} 
  \caption{The $x$-dependence of the mean squared transverse momentum $k_\perp$ of the unpolarized TMD $f_1$ for the leading Fock sector $|q\bar{q}\rangle$ and two Fock sectors $|q\bar{q}\rangle+|q\bar{q}g\rangle$ after averaging with the truncation parameters $N_{\text{max}}=\{12,14,16\}$, $K=15$ under the BLFQ framework.}
  \label{meansquaredkT}
\end{figure}

We consider the interpretation of the quark unpolarized TMD $f_1(x,k_\perp)$, which describes the probability of having an unpolarized active quark with the longitudinal momentum fraction $x$ and the relative transverse momentum $k_\perp$ in an unpolarized hadron \cite{Jaffe:1983hp}. We then define the $x$-dependent mean squared transverse momentum of the quark as
\begin{equation}
  \langle k_\perp^2\rangle_{f_1}(x)=\frac{\int\mathrm{d}^2k_\perp k_\perp^2 f_1(x,k_\perp)}{\int\mathrm{d}^2k_\perp  f_1(x,k_\perp)}.
\end{equation}
Figure \ref{meansquaredkT} shows the results for the $x$-dependence of the mean squared transverse momentum $\langle k_\perp^2\rangle_{f_1}(x)$ of the quark with different Fock components: $|q\bar{q}\rangle$ and $|q\bar{q}\rangle +|q\bar{q} g \rangle$. Both the distributions have their peaks at $x=0.5$. We observe that the $\langle k_\perp^2\rangle$ distribution from the leading Fock sector is symmetric about $x=0.5$. This occurs due to the same quark and antiquark masses in the pion.
Meanwhile, the $\langle k_\perp^2\rangle$ distribution from $|q\bar{q}\rangle+|q\bar{q}g\rangle$ is smaller than that from $|q\bar{q}\rangle$. This is attributed to the fact that the dynamical gluon in $|q\bar{q}g\rangle$ carries away a part of the quark (antiquark) momentum.

Although there is no probabilistic interpretation for twist-3 TMDs $e$ and $f^\perp$, we calculate their first and second $k_\perp$ moment defined as~\cite{Lorce:2014hxa}
\begin{equation}\label{moments}
  \langle k_\perp^n\rangle_{\rm TMD}=\frac{\int\mathrm{d}x\int\mathrm{d}^2{k}_\perp k_\perp^n {\rm TMD}}{\int\mathrm{d}x\int\mathrm{d}^2{k}_\perp  \mathrm{TMD}}.
\end{equation}
The corresponding results are summarized in Table \ref{kTmoment}. 
%
%
Our results are close to those from the light-front constituent model (LFCM) at a low scale $\mu_0\sim0.5$ GeV~\cite{Lorce:2016ugb}. Note that the latter model's results are based on the leading Fock component.

\begin{table}[h]
  \caption{$\langle k_\perp\rangle$ and $\langle k_\perp^2\rangle$ moments as defined in Eq.~\eqref{moments} 
  for the twist-2 TMD, $f_1(x,k_\perp)$, and the twist-3 TMDs, $e(x,k_\perp)$ and $f^\perp(x,k_\perp)$. The BLFQ results are all obtained by averaging with the truncation parameters $N_{\text{max}}=\{12,14,16\}$, $K=15$. The LFCM results are from Ref.~\cite{Lorce:2016ugb}. All are in units of GeV.
}
\vspace{0.15cm}
  \label{kTmoment}
  \centering
  \begin{tabular}{ccccc}
    \hline\hline
          & $\langle k_\perp\rangle_{\mathrm{BLFQ}}$ &$\langle k^2_\perp\rangle^{1/2}_{\mathrm{BLFQ}}$ &  $\langle k_\perp\rangle_{\mathrm{LFCM}}$ &$\langle k^2_\perp\rangle^{1/2}_{\mathrm{LFCM}}$\\        
    \hline 
        $f_1$ & 0.26 & 0.30 & 0.28 & 0.32\\   
        $e$ & 0.26 & 0.30 & 0.26 &  0.30\\       
        $f^\perp$ & 0.25 & 0.29 & 0.26 &  0.30\\     
\hline\hline
  \end{tabular}
\end{table}

\subsection{Twist-2 and twist-3 PDFs}
\begin{figure*}[h]
  \centering
    \includegraphics[width=0.298\textwidth]{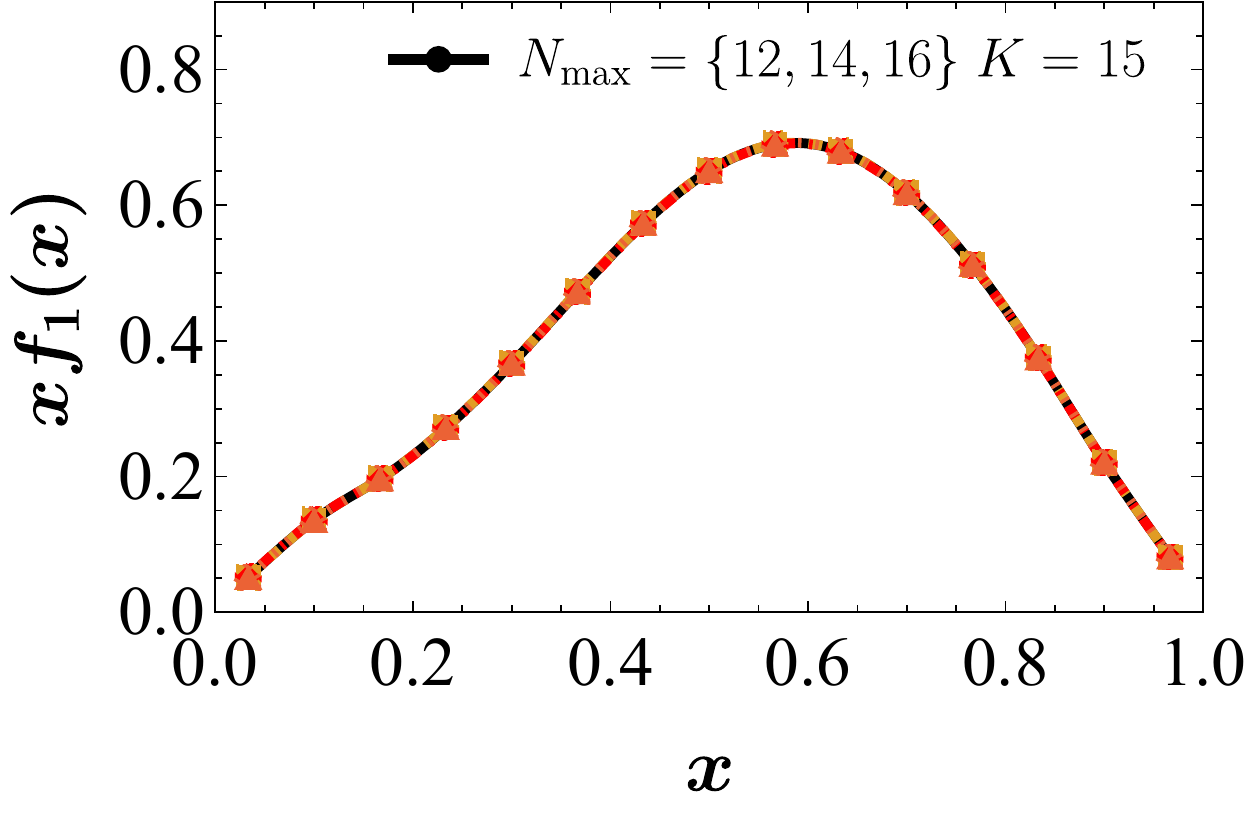}  
    \includegraphics[width=0.32\textwidth]{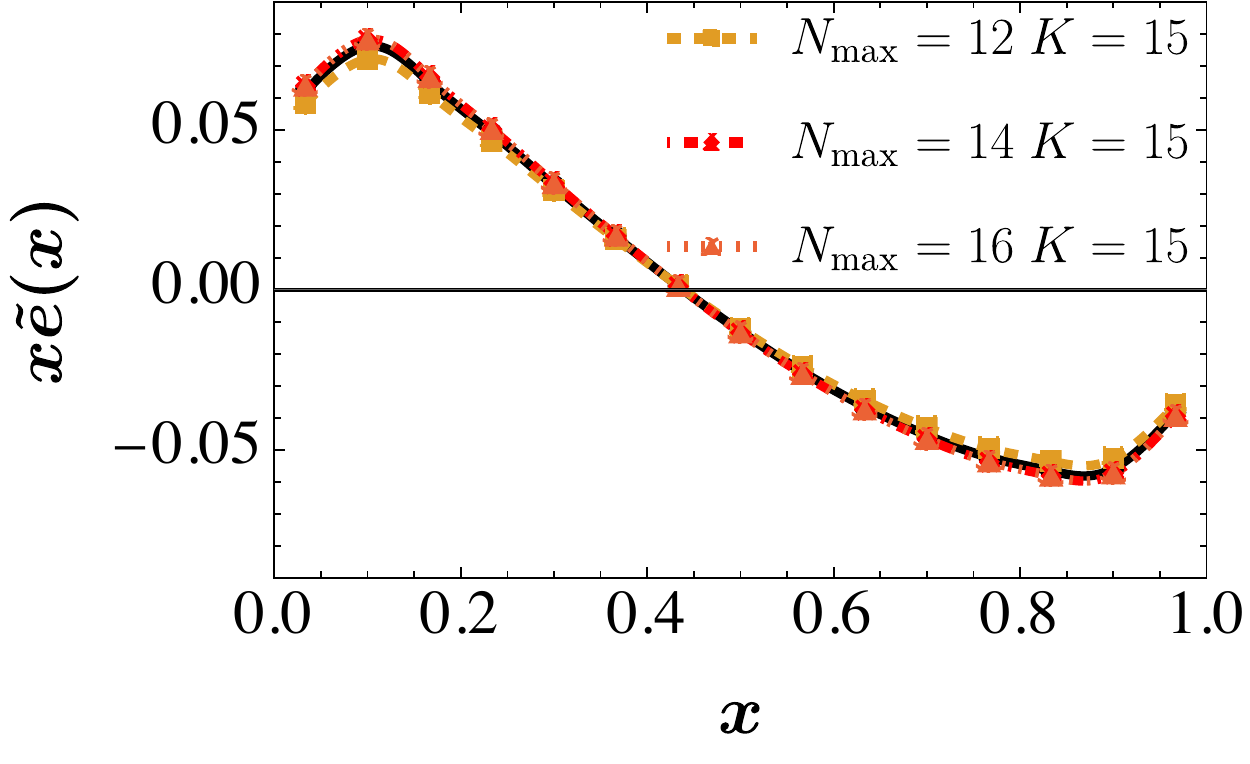}  
    \includegraphics[width=0.32\textwidth]{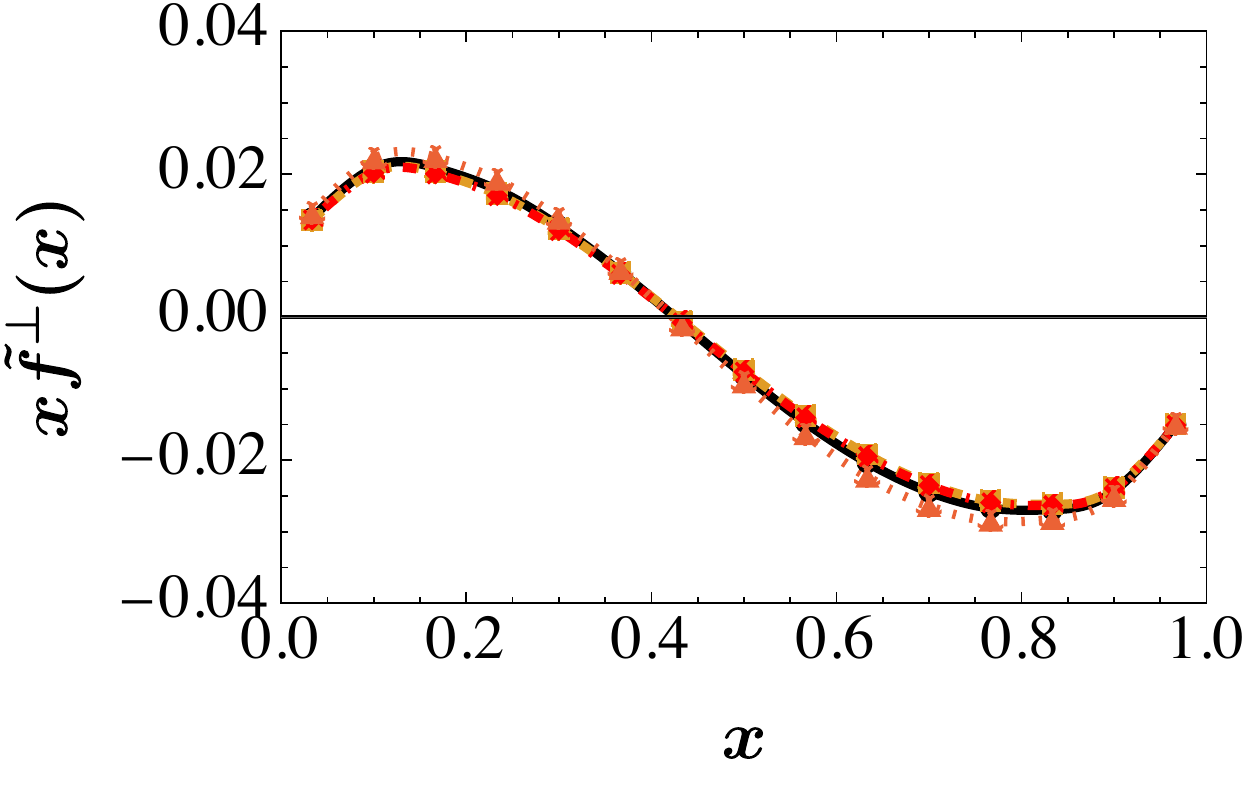}  
    \caption{The results for the twist-2 unpolarized PDF and the genuine twist-3 PDFs with the different $N_{\text{max}}$ under the BLFQ framework. }
    \label{PDFs}
\end{figure*}

In this subsection, we turn our attention to the integrated TMDs  namely the PDFs. 
Figure \ref{PDFs} shows the twist-2 PDF $xf_1(x)$, and the genuine twist-3 PDFs $x\tilde{e}(x)$ and $x\tilde{f}^\perp(x)$. Interestingly, and in contrast to our TMD results, the PDFs have no oscillations and are very stable for different $N_{\text{max}}$ truncations. 
Clearly, integrating the TMDs over $k_\perp$ eliminates the unphysical oscillations and yields smooth results for the PDFs. The twist-2 PDF $xf_1(x)$ shows a positive distribution, while the genuine twist-3 functions, $x\tilde{e}(x)$ and $x\tilde{f}^\perp(x)$, exhibit positive behavior when $x<0.5$, but they are negative above $x=0.5$. The magnitude of the twist-2 distribution is much higher than those of the twist-3 distributions. 
We also find that the integrated value of the distribution $x\tilde{e}(x)$ over $x$ vanishes showing that our BLFQ results properly satisfy the sum rule defined in Eq.~\eqref{SumRul_xetilde}. Note that this sum rule is derived from the EOM relations and thus, it must be independent of the system. The genuine twist-3 PDF $\tilde{e}(x)$ for the nucleon has been studied in Ref.~\cite{Pasquini:2018oyz}, where the authors considered $|qqq\rangle$ and $|qqqg\rangle$ Fock components but their distribution $\tilde{e}(x)$ does not satisfy the sum rule defined in Eq.~\eqref{SumRul_xetilde}. 
  In Ref.~\cite{Mukherjee:2009uy}, the author investigated the twist-3 PDF $e(x)$ and the genuine twist-3 PDF $\tilde{e}(x)$ for a dressed quark at one loop, where the distribution $\tilde{e}(x)$ satisfies the first-moment sum rule~(\ref{SumRul_xetilde}).
Using the sum rule defined in Eq.~(\ref{scalarFF}), we compute the pion scalar form factor from our BLFQ results, $\sigma_\pi=13.44$ GeV.


\begin{figure}[h]
  \centering
    \includegraphics[width=0.3\textwidth]{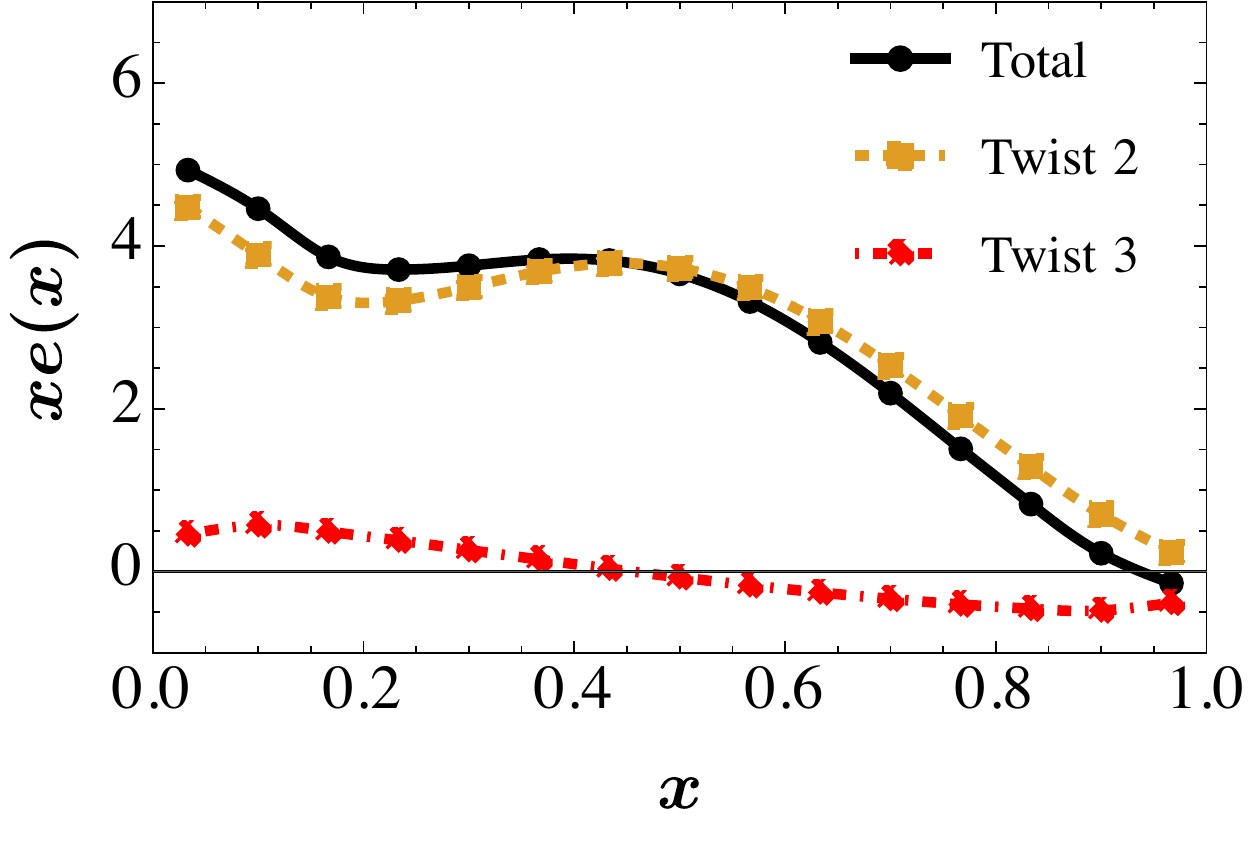}  
    \includegraphics[width=0.31\textwidth]{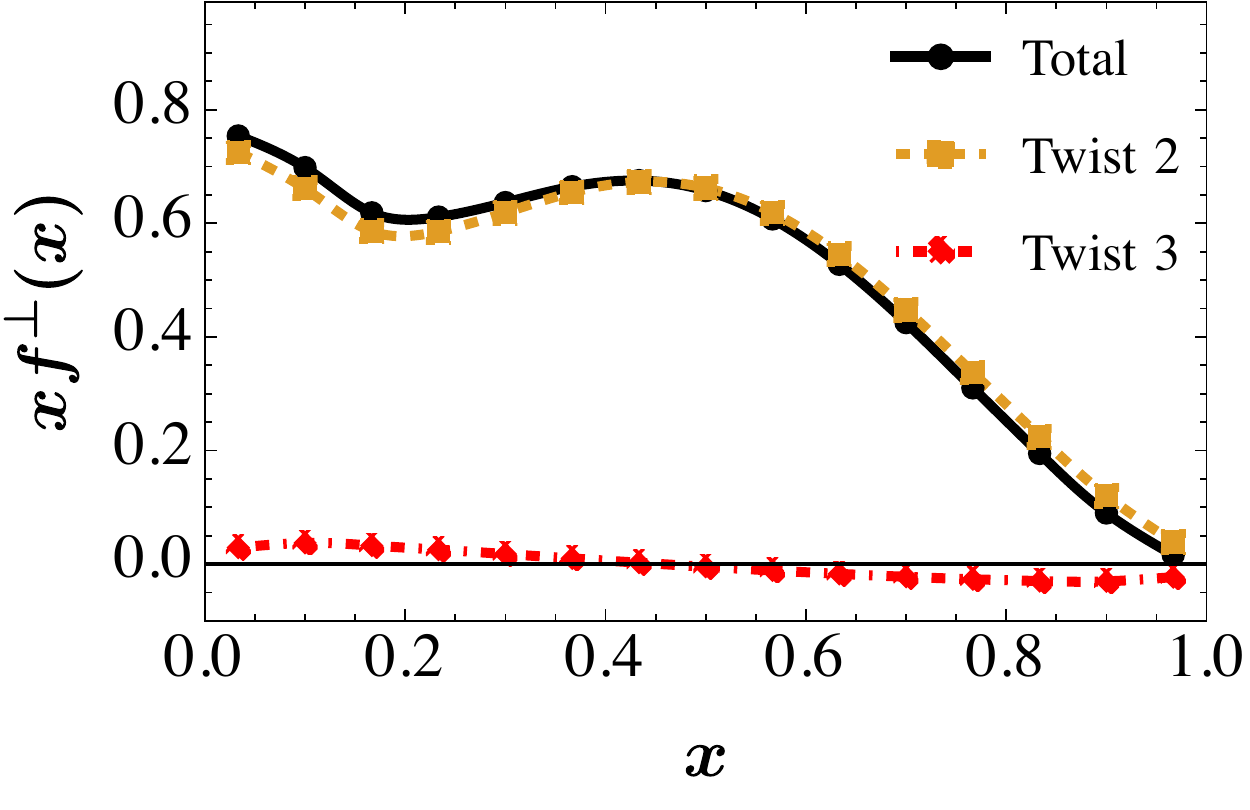} 
    \caption{The results for the total twist-3 PDFs, the genuine twist-3 PDFs, and the contributions from the twist-2 PDFs after averaging with the truncation parameters $N_{\text{max}}=\{12,14,16\}$, $K=15$ under the BLFQ framework.}\label{tw-3PDFs}
\end{figure}

Figure \ref{tw-3PDFs} shows the total twist-3 PDFs and their decompositions in terms of twist-2 and genuine twist-3 PDFs. We observe that the twist-2 terms and the genuine twist-3 PDFs have distinctly different $x$-dependence behaviors. The former strongly dominates the total twist-3 PDFs. 

\begin{figure*}[h]
  \centering
    \includegraphics[width=0.3\textwidth]{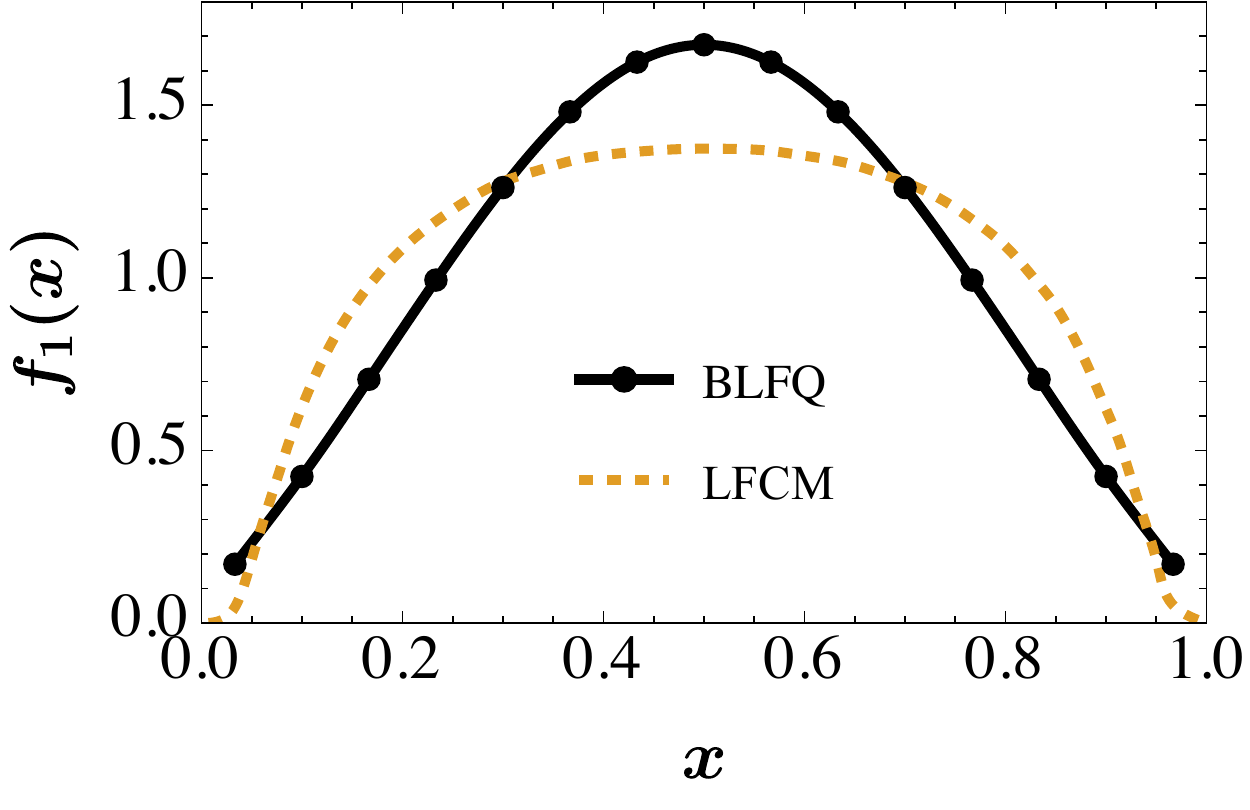}  
    \includegraphics[width=0.3\textwidth]{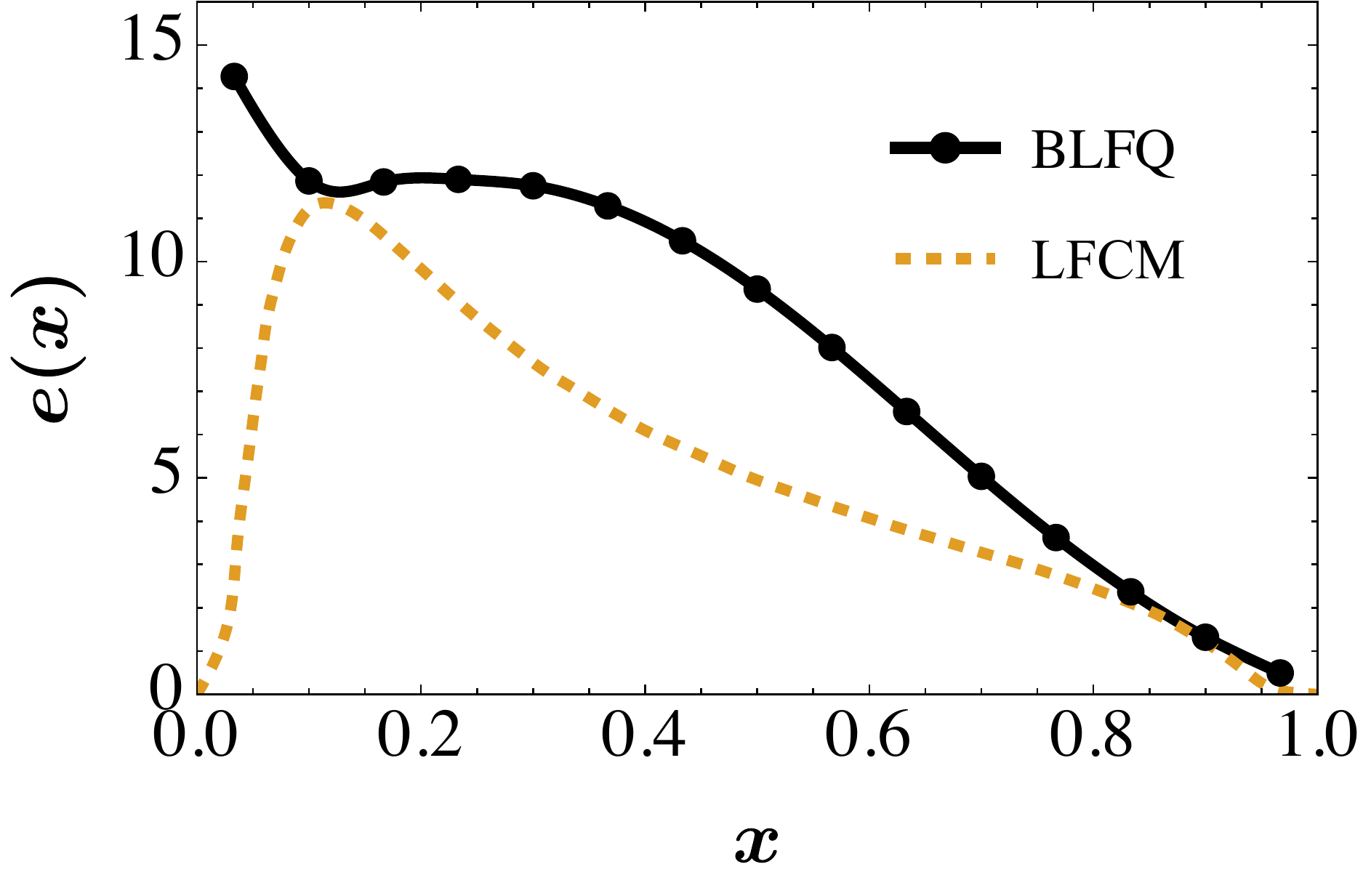}  
    \includegraphics[width=0.3\textwidth]{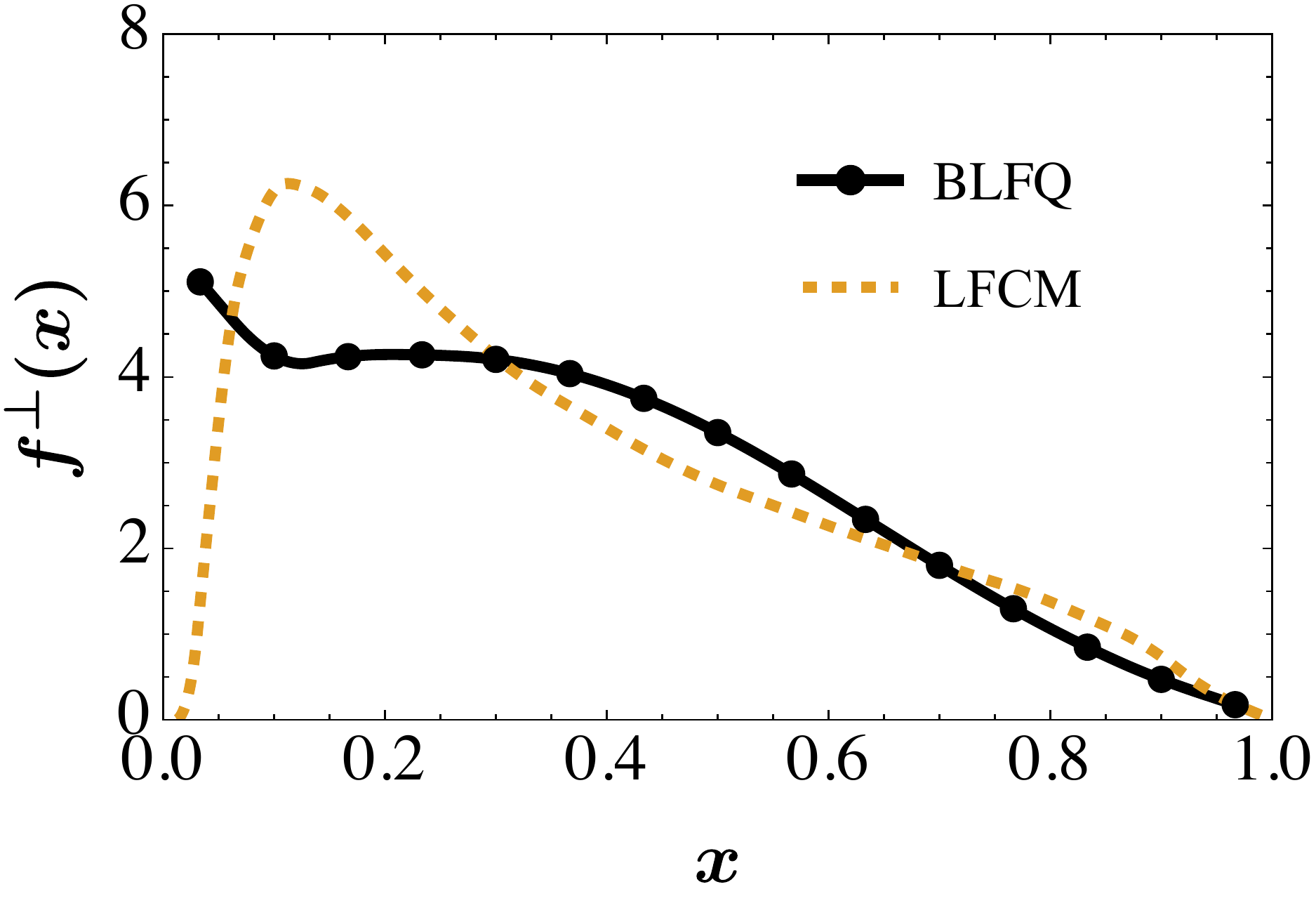}  
    \caption{Comparison of the unpolarized PDFs $f_1(x)$, $e(x)$ and $f^\perp(x)$ of the LFCM~\cite{Lorce:2016ugb} and the results after normalization to the leading Fork state $|q\bar{q}\rangle$ and after averaging $N_{\text{max}}=\{12,14,16\}$ of the BLFQ framework. }
    \label{comp}
\end{figure*}
Figure \ref{comp} compares our integrated TMDs $f_1(x)$, $e(x)$, and $f^\perp(x)$ computed from only contributions of the leading Fock sector $|q\bar{q}\rangle$ with the LFCM's results~\cite{Lorce:2016ugb}, which demonstrates the pion structure based on a two-body light-front Fock state $|q\bar{q}\rangle$. For $f_1(x)$, both models produce symmetric distributions about $x=0.5$ due to equal quark and antiquark mass. However, the $f_1(x)$ in the LFCM is wider, while our BLFQ result is narrower and steeper. The PDF $e(x)$ in the LFCM rises rapidly from zero in the small $x$ region and decreases quickly in the large $x$ region. Meanwhile, the BLFQ result shows a different trend for $e(x)$ dropping to a plateau in the small $x$ region and then continuing to drop to zero. We notice that $f^\perp (x)$ exhibits a somewhat similar trend as observed for $e(x)$. At large-$x$, both the LFCM and the BLFQ framework provide qualitatively similar behavior for $f^\perp (x)$. 

\section{Summary\label{Sec6}}
In this work, we calculated the twist-2 TMD $f_1(x,k_\perp)$ and the genuine twist-3 TMDs $\tilde{e}(x,k_\perp)$ and $\tilde{f}^\perp(x,k_\perp)$ of the pion at the model scale from its LFWFs within the basis light-front quantization framework. These wave functions have been obtained from the eigenvectors of the light-front QCD Hamiltonian in the light-cone gauge for the light mesons by considering them within $|q\bar{q}\rangle$ and $|q\bar{q}g\rangle$ Fock spaces, together with a 3D confinement in the leading Fock sector.
We employed the model-independent EOM relations to obtain the total twist-3 TMDs $e(x,k_\perp)$ and $f^\perp(x,k_\perp)$. 
In this study, the gauge link has been set to unity, which leaves us only the T-even quark TMDs. We also adopted an averaging procedure for reducing finite basis artifacts of our BLFQ results for the TMDs.
The integrated TMDs exhibit good consistency between computations with different basis truncations. Our BLFQ results satisfy the momentum sum rule related to $e(x)$. For the total twist-3 PDFs ${e}(x)$ and ${f}^\perp(x)$, we found that the contributions from the genuine twist-3 PDFs have a distinctive $x$-dependence from those of the twist-2 PDF. 
Due to imprecise knowledge of the nonperturbative Sudakov-like form factors and complexity of the evolution of high-twist TMDs~\cite{Rodini:2022wki} at present, it is not feasible yet to evolve our twist-3 TMDs in order to employ them for predicting experimental observables, such as spin asymmetries~\cite{Bacchetta:2019qkv}.


We compared our results from the leading Fock sector $|q\bar{q}\rangle$ with the LFCM. Our twist-3 PDFs show a different trend at small $x$ from the prediction of the LFCM, while at large $x$ both approaches provide similar qualitative behaviors. Considering that the genuine twist-3 TMDs and the T-odd TMDs with gauge links have similar operator structures, the results of this work are encouraging for an extension to the T-odd TMDs. Future studies will focus on the Boer-Mulders TMD $h_1^\perp$ of pions and its application to spin-asymmetries in Drell-Yan processes. 
In addition, a detailed analysis of the pion's gluon TMDs in the BLFQ will be reported in a future study.


We believe that these results including twist-3 TMDs provide predictions for future experiments, such as precise measurements of the pion-nucleus induced Drell-Yan processes and future measurements of the 3D structure of the pion at the EIC and at the EicC~\cite{Chavez:2021koz}, which will refine our 3D knowledge of the pion and its internal correlations based on QCD.


\section*{Acknowledgements}
We thank Siqi Xu, Jiatong Wu, Tiancai Peng and Zhe Liu for many helpful discussions. 
J. L. is supported by Special Research Assistant Funding Project, Chinese Academy of Sciences.
C. M. is supported by new faculty start up funding the Institute of Modern Physics, Chinese Academy of Sciences, Grants No. E129952YR0. C. M. also thanks the Chinese Academy of Sciences Presidents International Fellowship Initiative for the support via Grants No. 2021PM0023. 
X. Z. is supported by new faculty startup funding by the Institute of Modern Physics, Chinese Academy of Sciences, by Key Research Program of Frontier Sciences, Chinese Academy of Sciences, Grant No. ZDB-SLY-7020, by the Natural Science Foundation of Gansu Province, China, Grant No. 20JR10RA067, by the Foundation for Key Talents of Gansu Province, by the Central Funds Guiding the Local Science and Technology Development of Gansu Province, Grant No. 22ZY1QA006, and by the Strategic Priority Research Program of the Chinese Academy of Sciences, Grant No. XDB34000000. 
J. P. V. is supported by the Department of Energy under Grants No. DE-FG02-87ER40371, No. DE-SC0018223 (SciDAC4/NUCLEI) and DE-SC0023495 (SciDAC5/NUCLEI). A portion of the computational resources were also provided by Taiyuan Advanced Computing Center.

\appendix
\section{Derivation of the EOM relations}
In this appendix, we show the details of the derivation of the EOM relations given in Eqs.~(\ref{EOMeom1}) and (\ref{EOMeom2}). In Ref.~\cite{Lorce:2014hxa}, the authors derived EOM relations in a free quark model without considering the gauge link. Here, we also ignore the gauge links, however, our bound state contains a dynamical gluon.

For an arbitrary Dirac matrix $\Gamma$, we insert the EOM in the unintegrated correlation function, and we obtain
\begin{align}
    0&=\int\frac{d^4 z}{(2\pi)^4}e^{ik\cdot z}\langle P|\bar{\psi}(0)\Gamma(i\slashed{D}(z)-m_q)\psi(z)|P\rangle\nonumber\\
    &=\int\frac{d^4 z}{(2\pi)^4}e^{ik\cdot z}\langle P|\bar{\psi}(0)\Gamma[\slashed{k}+g\slashed{A}(z)-m_q]\psi(z)|P\rangle\label{EOM11},
\end{align} 
where the first term is derived from the surface integral. Adding this formula and its complex conjugate yields 
\begin{align}
      0=\int\frac{d^4 z}{(2\pi)^4}e^{ik\cdot z}\langle P|&\bar{\psi}(0)\{[\slashed{k}+g\slashed{A}(0)-m_q]\bar{\Gamma}\nonumber\\
      +&\Gamma[\slashed{k}+g\slashed{A}(z)-m_q]\}\psi(z)|P\rangle\label{uninteg},
\end{align} 
where $\bar{\Gamma}=\gamma^0\Gamma\gamma^0$. When $\Gamma=\gamma^+$, we integrate the unintegrated correlator of Eq. (\ref{uninteg}) over the light-front energy component $k^-$ and the light-front time component $z^+$ in the light-cone gauge $A^+=0$, and we obtain
\begin{align}
  0=\int&\frac{d^3 z}{(2\pi)^4}e^{ik\cdot z}\langle P|\bar{\psi}(0)[2k^+-2m_q\gamma^+\nonumber\\
  &-ig(A^j(z)-A^j(0))\sigma^{j+}]\psi(z)|P\rangle|_{z^+=0}.
\end{align}
From the above equation, we can find three kinds of gamma matrix structures, $\mathbbm{1}$, $\gamma^+$ and $\sigma^{j+}$. The three gamma structures, by definitions in Eqs.~(\ref{tw-2TMDs1}) and (\ref{tw-3TMDs-e}), correspond to the twist-3 TMD, $e(x,k_\perp)$, the twist-2 TMD, $f_1(x,k_\perp)$, and the genuine twist-3 TMD, $\tilde{e}(x,k_\perp)$, respectively. Arranging the formula, we get the EOM relation,
\begin{align}
  xe(x,k_\perp)&=x\tilde{e}(x,k_\perp)+\frac{m_q}{M}f_1(x,k_\perp),\label{EOM1}
\end{align}
where the genuine twist-3 TMD is defined as follows
\begin{align}
      &\tilde{e}(x,k_\perp)=\frac{i}{Mx}\frac{g}{2}\int\frac{\mathrm{d}z^-\mathrm{d}^2z_\perp}{2(2\pi)^3}e^{ikz}\nonumber\\
      &\times\langle P|\bar{\psi}_+(0)\sigma^{j+}[A_\perp^j(z)-A^j_\perp(0)]\psi_+(z)|P\rangle|_{z^+=0}.
\end{align}
We can also get a similar equation by setting $\Gamma=-i\sigma^{j+}$ in Eq.~(\ref{uninteg}) for the another T-even twist-3 TMD $\tilde{f}^\perp$ of spin-0 hadrons. 

The above derivation does not take into account the contributions from gauge links, but future inclusion of the gauge link is feasible yet will leave the EOM augmented with additional singular terms~\cite{Efremov:2002qh,Lorce:2014hxa,Pasquini:2018oyz}.


\biboptions{sort&compress}
\bibliographystyle{elsarticle-num}
\bibliography{PionTMDRef.bib}

\begin{thebibliography}{100}
\expandafter\ifx\csname url\endcsname\relax
  \def\url#1{\texttt{#1}}\fi
\expandafter\ifx\csname urlprefix\endcsname\relax\def\urlprefix{URL }\fi
\expandafter\ifx\csname href\endcsname\relax
  \def\href#1#2{#2} \def\path#1{#1}\fi

\bibitem{Accardi:2012qut}
A.~Accardi, et~al., {Electron Ion Collider: The Next QCD Frontier}:
  {Understanding the glue that binds us all}, Eur. Phys. J. A 52~(9) (2016)
  268.
\newblock \href {http://arxiv.org/abs/1212.1701} {\path{arXiv:1212.1701}},
  \href {https://doi.org/10.1140/epja/i2016-16268-9}
  {\path{doi:10.1140/epja/i2016-16268-9}}.

\bibitem{Bacchetta:2006tn}
A.~Bacchetta, M.~Diehl, K.~Goeke, A.~Metz, P.~J. Mulders, M.~Schlegel,
  {Semi-inclusive deep inelastic scattering at small transverse momentum}, JHEP
  02 (2007) 093.
\newblock \href {http://arxiv.org/abs/hep-ph/0611265}
  {\path{arXiv:hep-ph/0611265}}, \href
  {https://doi.org/10.1088/1126-6708/2007/02/093}
  {\path{doi:10.1088/1126-6708/2007/02/093}}.

\bibitem{Belitsky:2002sm}
A.~V. Belitsky, X.~Ji, F.~Yuan, {Final state interactions and gauge invariant
  parton distributions}, Nucl. Phys. B 656 (2003) 165--198.
\newblock \href {http://arxiv.org/abs/hep-ph/0208038}
  {\path{arXiv:hep-ph/0208038}}, \href
  {https://doi.org/10.1016/S0550-3213(03)00121-4}
  {\path{doi:10.1016/S0550-3213(03)00121-4}}.

\bibitem{Lai:2010vv}
H.-L. Lai, M.~Guzzi, J.~Huston, Z.~Li, P.~M. Nadolsky, J.~Pumplin, C.~P. Yuan,
  {New parton distributions for collider physics}, Phys. Rev. D 82 (2010)
  074024.
\newblock \href {http://arxiv.org/abs/1007.2241} {\path{arXiv:1007.2241}},
  \href {https://doi.org/10.1103/PhysRevD.82.074024}
  {\path{doi:10.1103/PhysRevD.82.074024}}.

\bibitem{Pumplin:2002vw}
J.~Pumplin, D.~R. Stump, J.~Huston, H.~L. Lai, P.~M. Nadolsky, W.~K. Tung, {New
  generation of parton distributions with uncertainties from global QCD
  analysis}, JHEP 07 (2002) 012.
\newblock \href {http://arxiv.org/abs/hep-ph/0201195}
  {\path{arXiv:hep-ph/0201195}}, \href
  {https://doi.org/10.1088/1126-6708/2002/07/012}
  {\path{doi:10.1088/1126-6708/2002/07/012}}.

\bibitem{Diehl:2003ny}
M.~Diehl, {Generalized parton distributions}, Phys. Rept. 388 (2003) 41--277.
\newblock \href {http://arxiv.org/abs/hep-ph/0307382}
  {\path{arXiv:hep-ph/0307382}}, \href
  {https://doi.org/10.1016/j.physrep.2003.08.002}
  {\path{doi:10.1016/j.physrep.2003.08.002}}.

\bibitem{Collins:2011zzd}
J.~Collins, {Foundations of perturbative QCD}, Vol.~32, Cambridge University
  Press, 2013.

\bibitem{Diehl:2011yj}
M.~Diehl, D.~Ostermeier, A.~Schafer, {Elements of a theory for multiparton
  interactions in QCD}, JHEP 03 (2012) 089, [Erratum: JHEP 03, 001 (2016)].
\newblock \href {http://arxiv.org/abs/1111.0910} {\path{arXiv:1111.0910}},
  \href {https://doi.org/10.1007/JHEP03(2012)089}
  {\path{doi:10.1007/JHEP03(2012)089}}.

\bibitem{Collins:1996fb}
J.~C. Collins, L.~Frankfurt, M.~Strikman, {Factorization for hard exclusive
  electroproduction of mesons in QCD}, Phys. Rev. D 56 (1997) 2982--3006.
\newblock \href {http://arxiv.org/abs/hep-ph/9611433}
  {\path{arXiv:hep-ph/9611433}}, \href
  {https://doi.org/10.1103/PhysRevD.56.2982}
  {\path{doi:10.1103/PhysRevD.56.2982}}.

\bibitem{Rogers:2015sqa}
T.~C. Rogers, {An overview of transverse-momentum\textendash{}dependent
  factorization and evolution}, Eur. Phys. J. A 52~(6) (2016) 153.
\newblock \href {http://arxiv.org/abs/1509.04766} {\path{arXiv:1509.04766}},
  \href {https://doi.org/10.1140/epja/i2016-16153-7}
  {\path{doi:10.1140/epja/i2016-16153-7}}.

\bibitem{Sterman:1995fz}
G.~F. Sterman, {Partons, factorization and resummation, TASI 95}, in:
  {Theoretical Advanced Study Institute in Elementary Particle Physics (TASI
  95): QCD and Beyond}, 1995, pp. 327--408.
\newblock \href {http://arxiv.org/abs/hep-ph/9606312}
  {\path{arXiv:hep-ph/9606312}}.

\bibitem{Collins:1985ue}
J.~C. Collins, D.~E. Soper, G.~F. Sterman, {Factorization for Short Distance
  Hadron - Hadron Scattering}, Nucl. Phys. B 261 (1985) 104--142.
\newblock \href {https://doi.org/10.1016/0550-3213(85)90565-6}
  {\path{doi:10.1016/0550-3213(85)90565-6}}.

\bibitem{Collins:1987pm}
J.~C. Collins, D.~E. Soper, {The Theorems of Perturbative QCD}, Ann. Rev. Nucl.
  Part. Sci. 37 (1987) 383--409.
\newblock \href {https://doi.org/10.1146/annurev.ns.37.120187.002123}
  {\path{doi:10.1146/annurev.ns.37.120187.002123}}.

\bibitem{Collins:1998rz}
J.~C. Collins, {Hard scattering factorization with heavy quarks: A General
  treatment}, Phys. Rev. D 58 (1998) 094002.
\newblock \href {http://arxiv.org/abs/hep-ph/9806259}
  {\path{arXiv:hep-ph/9806259}}, \href
  {https://doi.org/10.1103/PhysRevD.58.094002}
  {\path{doi:10.1103/PhysRevD.58.094002}}.

\bibitem{Chernyak:1983ej}
V.~L. Chernyak, A.~R. Zhitnitsky, {Asymptotic Behavior of Exclusive Processes
  in QCD}, Phys. Rept. 112 (1984) 173.
\newblock \href {https://doi.org/10.1016/0370-1573(84)90126-1}
  {\path{doi:10.1016/0370-1573(84)90126-1}}.

\bibitem{Muller:1994ses}
D.~M\"uller, D.~Robaschik, B.~Geyer, F.~M. Dittes, J.~Ho\v{r}ej\v{s}i, {Wave
  functions, evolution equations and evolution kernels from light ray operators
  of QCD}, Fortsch. Phys. 42 (1994) 101--141.
\newblock \href {http://arxiv.org/abs/hep-ph/9812448}
  {\path{arXiv:hep-ph/9812448}}, \href
  {https://doi.org/10.1002/prop.2190420202}
  {\path{doi:10.1002/prop.2190420202}}.

\bibitem{Lampe:1998eu}
B.~Lampe, E.~Reya, {Spin physics and polarized structure functions}, Phys.
  Rept. 332 (2000) 1--163.
\newblock \href {http://arxiv.org/abs/hep-ph/9810270}
  {\path{arXiv:hep-ph/9810270}}, \href
  {https://doi.org/10.1016/S0370-1573(99)00100-3}
  {\path{doi:10.1016/S0370-1573(99)00100-3}}.

\bibitem{Shuryak:1980tp}
E.~V. Shuryak, {Quantum Chromodynamics and the Theory of Superdense Matter},
  Phys. Rept. 61 (1980) 71--158.
\newblock \href {https://doi.org/10.1016/0370-1573(80)90105-2}
  {\path{doi:10.1016/0370-1573(80)90105-2}}.

\bibitem{Hen:2016kwk}
O.~Hen, G.~A. Miller, E.~Piasetzky, L.~B. Weinstein, {Nucleon-Nucleon
  Correlations, Short-lived Excitations, and the Quarks Within}, Rev. Mod.
  Phys. 89~(4) (2017) 045002.
\newblock \href {http://arxiv.org/abs/1611.09748} {\path{arXiv:1611.09748}},
  \href {https://doi.org/10.1103/RevModPhys.89.045002}
  {\path{doi:10.1103/RevModPhys.89.045002}}.

\bibitem{NNPDF:2017mvq}
R.~D. Ball, et~al., {Parton distributions from high-precision collider data},
  Eur. Phys. J. C 77~(10) (2017) 663.
\newblock \href {http://arxiv.org/abs/1706.00428} {\path{arXiv:1706.00428}},
  \href {https://doi.org/10.1140/epjc/s10052-017-5199-5}
  {\path{doi:10.1140/epjc/s10052-017-5199-5}}.

\bibitem{Jaffe:1983hp}
R.~L. Jaffe, {Parton Distribution Functions for Twist Four}, Nucl. Phys. B 229
  (1983) 205--230.
\newblock \href {https://doi.org/10.1016/0550-3213(83)90361-9}
  {\path{doi:10.1016/0550-3213(83)90361-9}}.

\bibitem{Ji:1998pc}
X.-D. Ji, {Off forward parton distributions}, J. Phys. G 24 (1998) 1181--1205.
\newblock \href {http://arxiv.org/abs/hep-ph/9807358}
  {\path{arXiv:hep-ph/9807358}}, \href
  {https://doi.org/10.1088/0954-3899/24/7/002}
  {\path{doi:10.1088/0954-3899/24/7/002}}.

\bibitem{Polyakov:1999gs}
M.~V. Polyakov, C.~Weiss, {Skewed and double distributions in pion and
  nucleon}, Phys. Rev. D 60 (1999) 114017.
\newblock \href {http://arxiv.org/abs/hep-ph/9902451}
  {\path{arXiv:hep-ph/9902451}}, \href
  {https://doi.org/10.1103/PhysRevD.60.114017}
  {\path{doi:10.1103/PhysRevD.60.114017}}.

\bibitem{Goeke:2001tz}
K.~Goeke, M.~V. Polyakov, M.~Vanderhaeghen, {Hard exclusive reactions and the
  structure of hadrons}, Prog. Part. Nucl. Phys. 47 (2001) 401--515.
\newblock \href {http://arxiv.org/abs/hep-ph/0106012}
  {\path{arXiv:hep-ph/0106012}}, \href
  {https://doi.org/10.1016/S0146-6410(01)00158-2}
  {\path{doi:10.1016/S0146-6410(01)00158-2}}.

\bibitem{Belitsky:2005qn}
A.~V. Belitsky, A.~V. Radyushkin, {Unraveling hadron structure with generalized
  parton distributions}, Phys. Rept. 418 (2005) 1--387.
\newblock \href {http://arxiv.org/abs/hep-ph/0504030}
  {\path{arXiv:hep-ph/0504030}}, \href
  {https://doi.org/10.1016/j.physrep.2005.06.002}
  {\path{doi:10.1016/j.physrep.2005.06.002}}.

\bibitem{Collins:2011ca}
J.~Collins, {New definition of TMD parton densities}, Int. J. Mod. Phys. Conf.
  Ser. 4 (2011) 85--96.
\newblock \href {http://arxiv.org/abs/1107.4123} {\path{arXiv:1107.4123}},
  \href {https://doi.org/10.1142/S2010194511001590}
  {\path{doi:10.1142/S2010194511001590}}.

\bibitem{Angeles-Martinez:2015sea}
R.~Angeles-Martinez, et~al., {Transverse Momentum Dependent (TMD) parton
  distribution functions: status and prospects}, Acta Phys. Polon. B 46~(12)
  (2015) 2501--2534.
\newblock \href {http://arxiv.org/abs/1507.05267} {\path{arXiv:1507.05267}},
  \href {https://doi.org/10.5506/APhysPolB.46.2501}
  {\path{doi:10.5506/APhysPolB.46.2501}}.

\bibitem{Diehl:2015uka}
M.~Diehl, {Introduction to GPDs and TMDs}, Eur. Phys. J. A 52~(6) (2016) 149.
\newblock \href {http://arxiv.org/abs/1512.01328} {\path{arXiv:1512.01328}},
  \href {https://doi.org/10.1140/epja/i2016-16149-3}
  {\path{doi:10.1140/epja/i2016-16149-3}}.

\bibitem{Bacchetta:2016ccz}
A.~Bacchetta, {Where do we stand with a 3-D picture of the proton?}, Eur. Phys.
  J. A 52~(6) (2016) 163.
\newblock \href {http://arxiv.org/abs/2107.06772} {\path{arXiv:2107.06772}},
  \href {https://doi.org/10.1140/epja/i2016-16163-5}
  {\path{doi:10.1140/epja/i2016-16163-5}}.

\bibitem{Burkardt:2000za}
M.~Burkardt, {Impact parameter dependent parton distributions and off forward
  parton distributions for zeta ---\ensuremath{>} 0}, Phys. Rev. D 62 (2000)
  071503, [Erratum: Phys.Rev.D 66, 119903 (2002)].
\newblock \href {http://arxiv.org/abs/hep-ph/0005108}
  {\path{arXiv:hep-ph/0005108}}, \href
  {https://doi.org/10.1103/PhysRevD.62.071503}
  {\path{doi:10.1103/PhysRevD.62.071503}}.

\bibitem{Burkardt:2002hr}
M.~Burkardt, {Impact parameter space interpretation for generalized parton
  distributions}, Int. J. Mod. Phys. A 18 (2003) 173--208.
\newblock \href {http://arxiv.org/abs/hep-ph/0207047}
  {\path{arXiv:hep-ph/0207047}}, \href
  {https://doi.org/10.1142/S0217751X03012370}
  {\path{doi:10.1142/S0217751X03012370}}.

\bibitem{Liu:2022fvl}
Y.~Liu, S.~Xu, C.~Mondal, X.~Zhao, J.~P. Vary, {Angular momentum and
  generalized parton distributions for the proton with basis light-front
  quantization}, Phys. Rev. D 105~(9) (2022) 094018.
\newblock \href {http://arxiv.org/abs/2202.00985} {\path{arXiv:2202.00985}},
  \href {https://doi.org/10.1103/PhysRevD.105.094018}
  {\path{doi:10.1103/PhysRevD.105.094018}}.

\bibitem{Ji:1996nm}
X.-D. Ji, {Deeply virtual Compton scattering}, Phys. Rev. D 55 (1997)
  7114--7125.
\newblock \href {http://arxiv.org/abs/hep-ph/9609381}
  {\path{arXiv:hep-ph/9609381}}, \href
  {https://doi.org/10.1103/PhysRevD.55.7114}
  {\path{doi:10.1103/PhysRevD.55.7114}}.

\bibitem{Favart:2015umi}
L.~Favart, M.~Guidal, T.~Horn, P.~Kroll, {Deeply Virtual Meson Production on
  the nucleon}, Eur. Phys. J. A 52~(6) (2016) 158.
\newblock \href {http://arxiv.org/abs/1511.04535} {\path{arXiv:1511.04535}},
  \href {https://doi.org/10.1140/epja/i2016-16158-2}
  {\path{doi:10.1140/epja/i2016-16158-2}}.

\bibitem{Boffi:2007yc}
S.~Boffi, B.~Pasquini, {Generalized parton distributions and the structure of
  the nucleon}, Riv. Nuovo Cim. 30~(9) (2007) 387--448.
\newblock \href {http://arxiv.org/abs/0711.2625} {\path{arXiv:0711.2625}},
  \href {https://doi.org/10.1393/ncr/i2007-10025-7}
  {\path{doi:10.1393/ncr/i2007-10025-7}}.

\bibitem{Boer:1997nt}
D.~Boer, P.~J. Mulders, {Time reversal odd distribution functions in
  leptoproduction}, Phys. Rev. D 57 (1998) 5780--5786.
\newblock \href {http://arxiv.org/abs/hep-ph/9711485}
  {\path{arXiv:hep-ph/9711485}}, \href
  {https://doi.org/10.1103/PhysRevD.57.5780}
  {\path{doi:10.1103/PhysRevD.57.5780}}.

\bibitem{Brodsky:2002rv}
S.~J. Brodsky, D.~S. Hwang, I.~Schmidt, {Initial state interactions and single
  spin asymmetries in Drell-Yan processes}, Nucl. Phys. B 642 (2002) 344--356.
\newblock \href {http://arxiv.org/abs/hep-ph/0206259}
  {\path{arXiv:hep-ph/0206259}}, \href
  {https://doi.org/10.1016/S0550-3213(02)00617-X}
  {\path{doi:10.1016/S0550-3213(02)00617-X}}.

\bibitem{Brodsky:2002cx}
S.~J. Brodsky, D.~S. Hwang, I.~Schmidt, {Final state interactions and single
  spin asymmetries in semiinclusive deep inelastic scattering}, Phys. Lett. B
  530 (2002) 99--107.
\newblock \href {http://arxiv.org/abs/hep-ph/0201296}
  {\path{arXiv:hep-ph/0201296}}, \href
  {https://doi.org/10.1016/S0370-2693(02)01320-5}
  {\path{doi:10.1016/S0370-2693(02)01320-5}}.

\bibitem{Mulders:1995dh}
P.~J. Mulders, R.~D. Tangerman, {The Complete tree level result up to order 1/Q
  for polarized deep inelastic leptoproduction}, Nucl. Phys. B 461 (1996)
  197--237, [Erratum: Nucl.Phys.B 484, 538--540 (1997)].
\newblock \href {http://arxiv.org/abs/hep-ph/9510301}
  {\path{arXiv:hep-ph/9510301}}, \href
  {https://doi.org/10.1016/0550-3213(95)00632-X}
  {\path{doi:10.1016/0550-3213(95)00632-X}}.

\bibitem{Goeke:2005hb}
K.~Goeke, A.~Metz, M.~Schlegel, {Parameterization of the quark-quark correlator
  of a spin-1/2 hadron}, Phys. Lett. B 618 (2005) 90--96.
\newblock \href {http://arxiv.org/abs/hep-ph/0504130}
  {\path{arXiv:hep-ph/0504130}}, \href
  {https://doi.org/10.1016/j.physletb.2005.05.037}
  {\path{doi:10.1016/j.physletb.2005.05.037}}.

\bibitem{Jaffe:1991ra}
R.~L. Jaffe, X.-D. Ji, {Chiral odd parton distributions and Drell-Yan
  processes}, Nucl. Phys. B 375 (1992) 527--560.
\newblock \href {https://doi.org/10.1016/0550-3213(92)90110-W}
  {\path{doi:10.1016/0550-3213(92)90110-W}}.

\bibitem{Jaffe:1991kp}
R.~L. Jaffe, X.-D. Ji, {Chiral odd parton distributions and polarized
  Drell-Yan}, Phys. Rev. Lett. 67 (1991) 552--555.
\newblock \href {https://doi.org/10.1103/PhysRevLett.67.552}
  {\path{doi:10.1103/PhysRevLett.67.552}}.

\bibitem{Efremov:2002qh}
A.~V. Efremov, P.~Schweitzer, {The Chirally odd twist 3 distribution e(a)(x)},
  JHEP 08 (2003) 006.
\newblock \href {http://arxiv.org/abs/hep-ph/0212044}
  {\path{arXiv:hep-ph/0212044}}, \href
  {https://doi.org/10.1088/1126-6708/2003/08/006}
  {\path{doi:10.1088/1126-6708/2003/08/006}}.

\bibitem{Burkardt:2008ps}
M.~Burkardt, {Transverse force on quarks in deep-inelastic scattering}, Phys.
  Rev. D 88 (2013) 114502.
\newblock \href {http://arxiv.org/abs/0810.3589} {\path{arXiv:0810.3589}},
  \href {https://doi.org/10.1103/PhysRevD.88.114502}
  {\path{doi:10.1103/PhysRevD.88.114502}}.

\bibitem{E12017}
H.~Avakian, et~al., Jlab experiment (2008).

\bibitem{E12-06-112}
H.~Avakian, et~al., Jlab experiment (2006).

\bibitem{E12-06-112B}
S.~Pisano, et~al., Jlab experiment (2014).

\bibitem{AbdulKhalek:2021gbh}
R.~Abdul~Khalek, et~al., {Science Requirements and Detector Concepts for the
  Electron-Ion Collider}: {EIC Yellow Report}, Nucl. Phys. A 1026 (2022)
  122447.
\newblock \href {http://arxiv.org/abs/2103.05419} {\path{arXiv:2103.05419}},
  \href {https://doi.org/10.1016/j.nuclphysa.2022.122447}
  {\path{doi:10.1016/j.nuclphysa.2022.122447}}.

\bibitem{AbdulKhalek:2022hcn}
R.~Abdul~Khalek, et~al., {Snowmass 2021 White Paper: Electron Ion Collider for
  High Energy Physics} (3 2022).
\newblock \href {http://arxiv.org/abs/2203.13199} {\path{arXiv:2203.13199}}.

\bibitem{Amoroso:2022eow}
S.~Amoroso, et~al., {Snowmass 2021 whitepaper: Proton structure at the
  precision frontier} (3 2022).
\newblock \href {http://arxiv.org/abs/2203.13923} {\path{arXiv:2203.13923}}.

\bibitem{Anderle:2021wcy}
D.~P. Anderle, et~al., {Electron-ion collider in China}, Front. Phys. (Beijing)
  16~(6) (2021) 64701.
\newblock \href {http://arxiv.org/abs/2102.09222} {\path{arXiv:2102.09222}},
  \href {https://doi.org/10.1007/s11467-021-1062-0}
  {\path{doi:10.1007/s11467-021-1062-0}}.

\bibitem{Boer:2011fh}
D.~Boer, et~al., {Gluons and the quark sea at high energies: Distributions,
  polarization, tomography} (8 2011).
\newblock \href {http://arxiv.org/abs/1108.1713} {\path{arXiv:1108.1713}}.

\bibitem{Jakob:1997wg}
R.~Jakob, P.~J. Mulders, J.~Rodrigues, {Modeling quark distribution and
  fragmentation functions}, Nucl. Phys. A 626 (1997) 937--965.
\newblock \href {http://arxiv.org/abs/hep-ph/9704335}
  {\path{arXiv:hep-ph/9704335}}, \href
  {https://doi.org/10.1016/S0375-9474(97)00588-5}
  {\path{doi:10.1016/S0375-9474(97)00588-5}}.

\bibitem{Lu:2012gu}
Z.~Lu, I.~Schmidt, {T-odd quark-gluon-quark correlation function in the diquark
  model}, Phys. Lett. B 712 (2012) 451--455.
\newblock \href {http://arxiv.org/abs/1202.0700} {\path{arXiv:1202.0700}},
  \href {https://doi.org/10.1016/j.physletb.2012.05.023}
  {\path{doi:10.1016/j.physletb.2012.05.023}}.

\bibitem{Mao:2013waa}
W.~Mao, Z.~Lu, {Beam spin asymmetries of charged and neutral pion production in
  semi-inclusive DIS}, Eur. Phys. J. C 73 (2013) 2557.
\newblock \href {http://arxiv.org/abs/1306.1004} {\path{arXiv:1306.1004}},
  \href {https://doi.org/10.1140/epjc/s10052-013-2557-9}
  {\path{doi:10.1140/epjc/s10052-013-2557-9}}.

\bibitem{Mao:2014aoa}
W.~Mao, Z.~Lu, B.-Q. Ma, {Transverse single-spin asymmetries of pion production
  in semi-inclusive DIS at subleading twist}, Phys. Rev. D 90~(1) (2014)
  014048.
\newblock \href {http://arxiv.org/abs/1405.3876} {\path{arXiv:1405.3876}},
  \href {https://doi.org/10.1103/PhysRevD.90.014048}
  {\path{doi:10.1103/PhysRevD.90.014048}}.

\bibitem{Signal:1996ct}
A.~I. Signal, {Calculations of higher twist distribution functions in the MIT
  bag model}, Nucl. Phys. B 497 (1997) 415--434.
\newblock \href {http://arxiv.org/abs/hep-ph/9610480}
  {\path{arXiv:hep-ph/9610480}}, \href
  {https://doi.org/10.1016/S0550-3213(97)00231-9}
  {\path{doi:10.1016/S0550-3213(97)00231-9}}.

\bibitem{Avakian:2010br}
H.~Avakian, A.~V. Efremov, P.~Schweitzer, F.~Yuan, {The transverse momentum
  dependent distribution functions in the bag model}, Phys. Rev. D 81 (2010)
  074035.
\newblock \href {http://arxiv.org/abs/1001.5467} {\path{arXiv:1001.5467}},
  \href {https://doi.org/10.1103/PhysRevD.81.074035}
  {\path{doi:10.1103/PhysRevD.81.074035}}.

\bibitem{Lorce:2014hxa}
C.~Lorc\'e, B.~Pasquini, P.~Schweitzer, {Unpolarized transverse momentum
  dependent parton distribution functions beyond leading twist in quark
  models}, JHEP 01 (2015) 103.
\newblock \href {http://arxiv.org/abs/1411.2550} {\path{arXiv:1411.2550}},
  \href {https://doi.org/10.1007/JHEP01(2015)103}
  {\path{doi:10.1007/JHEP01(2015)103}}.

\bibitem{Schweitzer:2003uy}
P.~Schweitzer, {The Chirally odd twist three distribution function e**alpha(x)
  in the chiral quark soliton model}, Phys. Rev. D 67 (2003) 114010.
\newblock \href {http://arxiv.org/abs/hep-ph/0303011}
  {\path{arXiv:hep-ph/0303011}}, \href
  {https://doi.org/10.1103/PhysRevD.67.114010}
  {\path{doi:10.1103/PhysRevD.67.114010}}.

\bibitem{Wakamatsu:2007nc}
M.~Wakamatsu, {Comparative analysis of the transversities and the
  longitudinally polarized distribution functions of the nucleon}, Phys. Lett.
  B 653 (2007) 398--403.
\newblock \href {http://arxiv.org/abs/0705.2917} {\path{arXiv:0705.2917}},
  \href {https://doi.org/10.1016/j.physletb.2007.08.013}
  {\path{doi:10.1016/j.physletb.2007.08.013}}.

\bibitem{Wakamatsu:2003uu}
M.~Wakamatsu, Y.~Ohnishi, {The Nonperturbative origin of delta function
  singularity in the chirally odd twist three distribution function e(x)},
  Phys. Rev. D 67 (2003) 114011.
\newblock \href {http://arxiv.org/abs/hep-ph/0303007}
  {\path{arXiv:hep-ph/0303007}}, \href
  {https://doi.org/10.1103/PhysRevD.67.114011}
  {\path{doi:10.1103/PhysRevD.67.114011}}.

\bibitem{Ohnishi:2003mf}
Y.~Ohnishi, M.~Wakamatsu, {pi N sigma term and chiral odd twist three
  distribution function e(x) of the nucleon in the chiral quark soliton model},
  Phys. Rev. D 69 (2004) 114002.
\newblock \href {http://arxiv.org/abs/hep-ph/0312044}
  {\path{arXiv:hep-ph/0312044}}, \href
  {https://doi.org/10.1103/PhysRevD.69.114002}
  {\path{doi:10.1103/PhysRevD.69.114002}}.

\bibitem{Cebulla:2007ej}
C.~Cebulla, J.~Ossmann, P.~Schweitzer, D.~Urbano, {The Twist-3 parton
  distribution function e**a(x) in large-N(c) chiral theory}, Acta Phys. Polon.
  B 39 (2008) 609--640.
\newblock \href {http://arxiv.org/abs/0710.3103} {\path{arXiv:0710.3103}}.

\bibitem{Balla:1997hf}
J.~Balla, M.~V. Polyakov, C.~Weiss, {Nucleon matrix elements of higher twist
  operators from the instanton vacuum}, Nucl. Phys. B 510 (1998) 327--364.
\newblock \href {http://arxiv.org/abs/hep-ph/9707515}
  {\path{arXiv:hep-ph/9707515}}, \href
  {https://doi.org/10.1016/S0550-3213(98)00638-5}
  {\path{doi:10.1016/S0550-3213(98)00638-5}}.

\bibitem{Dressler:1999hc}
B.~Dressler, M.~V. Polyakov, {On the twist - three contribution to h(L) in the
  instanton vacuum}, Phys. Rev. D 61 (2000) 097501.
\newblock \href {http://arxiv.org/abs/hep-ph/9912376}
  {\path{arXiv:hep-ph/9912376}}, \href
  {https://doi.org/10.1103/PhysRevD.61.097501}
  {\path{doi:10.1103/PhysRevD.61.097501}}.

\bibitem{Lorce:2016ugb}
C.~Lorc\'e, B.~Pasquini, P.~Schweitzer, {Transverse pion structure beyond
  leading twist in constituent models}, Eur. Phys. J. C 76~(7) (2016) 415.
\newblock \href {http://arxiv.org/abs/1605.00815} {\path{arXiv:1605.00815}},
  \href {https://doi.org/10.1140/epjc/s10052-016-4257-8}
  {\path{doi:10.1140/epjc/s10052-016-4257-8}}.

\bibitem{Pasquini:2018oyz}
B.~Pasquini, S.~Rodini, {The twist-three distribution $e^q(x,k_\perp)$ in a
  light-front model}, Phys. Lett. B 788 (2019) 414--424.
\newblock \href {http://arxiv.org/abs/1806.10932} {\path{arXiv:1806.10932}},
  \href {https://doi.org/10.1016/j.physletb.2018.11.033}
  {\path{doi:10.1016/j.physletb.2018.11.033}}.

\bibitem{Burkardt:2001iy}
M.~Burkardt, Y.~Koike, {Violation of sum rules for twist three parton
  distributions in QCD}, Nucl. Phys. B 632 (2002) 311--329.
\newblock \href {http://arxiv.org/abs/hep-ph/0111343}
  {\path{arXiv:hep-ph/0111343}}, \href
  {https://doi.org/10.1016/S0550-3213(02)00263-8}
  {\path{doi:10.1016/S0550-3213(02)00263-8}}.

\bibitem{Kundu:2001pk}
R.~Kundu, A.~Metz, {Higher twist and transverse momentum dependent parton
  distributions: A Light front Hamiltonian approach}, Phys. Rev. D 65 (2002)
  014009.
\newblock \href {http://arxiv.org/abs/hep-ph/0107073}
  {\path{arXiv:hep-ph/0107073}}, \href
  {https://doi.org/10.1103/PhysRevD.65.014009}
  {\path{doi:10.1103/PhysRevD.65.014009}}.

\bibitem{Mukherjee:2009uy}
A.~Mukherjee, {Twist Three Distribution e(x): Sum Rules and Equation of Motion
  Relations}, Phys. Lett. B 687 (2010) 180--183.
\newblock \href {http://arxiv.org/abs/0912.1446} {\path{arXiv:0912.1446}},
  \href {https://doi.org/10.1016/j.physletb.2010.03.023}
  {\path{doi:10.1016/j.physletb.2010.03.023}}.

\bibitem{Accardi:2009au}
A.~Accardi, A.~Bacchetta, W.~Melnitchouk, M.~Schlegel, {What can break the
  Wandzura-Wilczek relation?}, JHEP 11 (2009) 093.
\newblock \href {http://arxiv.org/abs/0907.2942} {\path{arXiv:0907.2942}},
  \href {https://doi.org/10.1088/1126-6708/2009/11/093}
  {\path{doi:10.1088/1126-6708/2009/11/093}}.

\bibitem{Ydrefors:2023src}
E.~Ydrefors, W.~de~Paula, T.~Frederico, G.~Salm\`e, {Unpolarized
  transverse-momentum dependent distribution functions of a quark in a pion
  with Minkowskian dynamics} (1 2023).
\newblock \href {http://arxiv.org/abs/2301.11599} {\path{arXiv:2301.11599}}.

\bibitem{Cerutti:2022lmb}
M.~Cerutti, L.~Rossi, S.~Venturini, A.~Bacchetta, V.~Bertone, C.~Bissolotti,
  M.~Radici, {Extraction of pion transverse momentum distributions from
  Drell-Yan data}, Phys. Rev. D 107~(1) (2023) 014014.
\newblock \href {http://arxiv.org/abs/2210.01733} {\path{arXiv:2210.01733}},
  \href {https://doi.org/10.1103/PhysRevD.107.014014}
  {\path{doi:10.1103/PhysRevD.107.014014}}.

\bibitem{Barry:2023qqh}
P.~C. Barry, L.~Gamberg, W.~Melnitchouk, E.~Moffat, D.~Pitonyak, A.~Prokudin,
  N.~Sato, {Tomography of pions and protons via transverse momentum dependent
  distributions} (2 2023).
\newblock \href {http://arxiv.org/abs/2302.01192} {\path{arXiv:2302.01192}}.

\bibitem{Vary:2009gt}
J.~P. Vary, H.~Honkanen, J.~Li, P.~Maris, S.~J. Brodsky, A.~Harindranath, G.~F.
  de~Teramond, P.~Sternberg, E.~G. Ng, C.~Yang, {Hamiltonian light-front field
  theory in a basis function approach}, Phys. Rev. C 81 (2010) 035205.
\newblock \href {http://arxiv.org/abs/0905.1411} {\path{arXiv:0905.1411}},
  \href {https://doi.org/10.1103/PhysRevC.81.035205}
  {\path{doi:10.1103/PhysRevC.81.035205}}.

\bibitem{Zhao:2014xaa}
X.~Zhao, H.~Honkanen, P.~Maris, J.~P. Vary, S.~J. Brodsky, {Electron g-2 in
  Light-Front Quantization}, Phys. Lett. B 737 (2014) 65--69.
\newblock \href {http://arxiv.org/abs/1402.4195} {\path{arXiv:1402.4195}},
  \href {https://doi.org/10.1016/j.physletb.2014.08.020}
  {\path{doi:10.1016/j.physletb.2014.08.020}}.

\bibitem{Wiecki:2014ola}
P.~Wiecki, Y.~Li, X.~Zhao, P.~Maris, J.~P. Vary, {Basis Light-Front
  Quantization Approach to Positronium}, Phys. Rev. D 91~(10) (2015) 105009.
\newblock \href {http://arxiv.org/abs/1404.6234} {\path{arXiv:1404.6234}},
  \href {https://doi.org/10.1103/PhysRevD.91.105009}
  {\path{doi:10.1103/PhysRevD.91.105009}}.

\bibitem{Li:2015zda}
Y.~Li, P.~Maris, X.~Zhao, J.~P. Vary, {Heavy Quarkonium in a Holographic
  Basis}, Phys. Lett. B 758 (2016) 118--124.
\newblock \href {http://arxiv.org/abs/1509.07212} {\path{arXiv:1509.07212}},
  \href {https://doi.org/10.1016/j.physletb.2016.04.065}
  {\path{doi:10.1016/j.physletb.2016.04.065}}.

\bibitem{Jia:2018ary}
S.~Jia, J.~P. Vary, {Basis light front quantization for the charged light
  mesons with color singlet Nambu\textendash{}Jona-Lasinio interactions}, Phys.
  Rev. C 99~(3) (2019) 035206.
\newblock \href {http://arxiv.org/abs/1811.08512} {\path{arXiv:1811.08512}},
  \href {https://doi.org/10.1103/PhysRevC.99.035206}
  {\path{doi:10.1103/PhysRevC.99.035206}}.

\bibitem{Qian:2020utg}
W.~Qian, S.~Jia, Y.~Li, J.~P. Vary, {Light mesons within the basis light-front
  quantization framework}, Phys. Rev. C 102~(5) (2020) 055207.
\newblock \href {http://arxiv.org/abs/2005.13806} {\path{arXiv:2005.13806}},
  \href {https://doi.org/10.1103/PhysRevC.102.055207}
  {\path{doi:10.1103/PhysRevC.102.055207}}.

\bibitem{Tang:2019gvn}
S.~Tang, Y.~Li, P.~Maris, J.~P. Vary, {Heavy-light mesons on the light front},
  Eur. Phys. J. C 80~(6) (2020) 522.
\newblock \href {http://arxiv.org/abs/1912.02088} {\path{arXiv:1912.02088}},
  \href {https://doi.org/10.1140/epjc/s10052-020-8081-9}
  {\path{doi:10.1140/epjc/s10052-020-8081-9}}.

\bibitem{Mondal:2019jdg}
C.~Mondal, S.~Xu, J.~Lan, X.~Zhao, Y.~Li, D.~Chakrabarti, J.~P. Vary, {Proton
  structure from a light-front Hamiltonian}, Phys. Rev. D 102~(1) (2020)
  016008.
\newblock \href {http://arxiv.org/abs/1911.10913} {\path{arXiv:1911.10913}},
  \href {https://doi.org/10.1103/PhysRevD.102.016008}
  {\path{doi:10.1103/PhysRevD.102.016008}}.

\bibitem{Xu:2021wwj}
S.~Xu, C.~Mondal, J.~Lan, X.~Zhao, Y.~Li, J.~P. Vary, {Nucleon structure from
  basis light-front quantization}, Phys. Rev. D 104~(9) (2021) 094036.
\newblock \href {http://arxiv.org/abs/2108.03909} {\path{arXiv:2108.03909}},
  \href {https://doi.org/10.1103/PhysRevD.104.094036}
  {\path{doi:10.1103/PhysRevD.104.094036}}.

\bibitem{Nair:2022evk}
S.~Nair, C.~Mondal, X.~Zhao, A.~Mukherjee, J.~P. Vary, {Basis light-front
  quantization approach to photon}, Phys. Lett. B 827 (2022) 137005.
\newblock \href {http://arxiv.org/abs/2201.12770} {\path{arXiv:2201.12770}},
  \href {https://doi.org/10.1016/j.physletb.2022.137005}
  {\path{doi:10.1016/j.physletb.2022.137005}}.

\bibitem{Lan:2021wok}
J.~Lan, K.~Fu, C.~Mondal, X.~Zhao, j.~P. Vary, {Light mesons with one dynamical
  gluon on the light front}, Phys. Lett. B 825 (2022) 136890.
\newblock \href {http://arxiv.org/abs/2106.04954} {\path{arXiv:2106.04954}},
  \href {https://doi.org/10.1016/j.physletb.2022.136890}
  {\path{doi:10.1016/j.physletb.2022.136890}}.

\bibitem{Adhikari:2021jrh}
L.~Adhikari, C.~Mondal, S.~Nair, S.~Xu, S.~Jia, X.~Zhao, J.~P. Vary,
  {Generalized parton distributions and spin structures of light mesons from a
  light-front Hamiltonian approach}, Phys. Rev. D 104~(11) (2021) 114019.
\newblock \href {http://arxiv.org/abs/2110.05048} {\path{arXiv:2110.05048}},
  \href {https://doi.org/10.1103/PhysRevD.104.114019}
  {\path{doi:10.1103/PhysRevD.104.114019}}.

\bibitem{Mondal:2021czk}
C.~Mondal, S.~Nair, S.~Jia, X.~Zhao, J.~P. Vary, {Pion to photon transition
  form factors with basis light-front quantization}, Phys. Rev. D 104~(9)
  (2021) 094034.
\newblock \href {http://arxiv.org/abs/2109.02279} {\path{arXiv:2109.02279}},
  \href {https://doi.org/10.1103/PhysRevD.104.094034}
  {\path{doi:10.1103/PhysRevD.104.094034}}.

\bibitem{Hu:2020arv}
Z.~Hu, S.~Xu, C.~Mondal, X.~Zhao, J.~P. Vary, {Transverse structure of electron
  in momentum space in basis light-front quantization}, Phys. Rev. D 103~(3)
  (2021) 036005.
\newblock \href {http://arxiv.org/abs/2010.12498} {\path{arXiv:2010.12498}},
  \href {https://doi.org/10.1103/PhysRevD.103.036005}
  {\path{doi:10.1103/PhysRevD.103.036005}}.

\bibitem{Lan:2019img}
J.~Lan, C.~Mondal, M.~Li, Y.~Li, S.~Tang, X.~Zhao, J.~P. Vary, {Parton
  Distribution Functions of Heavy Mesons on the Light Front}, Phys. Rev. D
  102~(1) (2020) 014020.
\newblock \href {http://arxiv.org/abs/1911.11676} {\path{arXiv:1911.11676}},
  \href {https://doi.org/10.1103/PhysRevD.102.014020}
  {\path{doi:10.1103/PhysRevD.102.014020}}.

\bibitem{Lan:2019rba}
J.~Lan, C.~Mondal, S.~Jia, X.~Zhao, J.~P. Vary, {Pion and kaon parton
  distribution functions from basis light front quantization and QCD
  evolution}, Phys. Rev. D 101~(3) (2020) 034024.
\newblock \href {http://arxiv.org/abs/1907.01509} {\path{arXiv:1907.01509}},
  \href {https://doi.org/10.1103/PhysRevD.101.034024}
  {\path{doi:10.1103/PhysRevD.101.034024}}.

\bibitem{Lan:2019vui}
J.~Lan, C.~Mondal, S.~Jia, X.~Zhao, J.~P. Vary, {Parton Distribution Functions
  from a Light Front Hamiltonian and QCD Evolution for Light Mesons}, Phys.
  Rev. Lett. 122~(17) (2019) 172001.
\newblock \href {http://arxiv.org/abs/1901.11430} {\path{arXiv:1901.11430}},
  \href {https://doi.org/10.1103/PhysRevLett.122.172001}
  {\path{doi:10.1103/PhysRevLett.122.172001}}.

\bibitem{Xu:2022abw}
S.~Xu, C.~Mondal, X.~Zhao, Y.~Li, J.~P. Vary, {Nucleon spin decomposition with
  one dynamical gluon} (9 2022).
\newblock \href {http://arxiv.org/abs/2209.08584} {\path{arXiv:2209.08584}}.

\bibitem{Peng:2022lte}
T.~Peng, Z.~Zhu, S.~Xu, X.~Liu, C.~Mondal, X.~Zhao, J.~P. Vary, {Basis
  light-front quantization approach to \ensuremath{\Lambda} and
  \ensuremath{\Lambda}c and their isospin triplet baryons}, Phys. Rev. D
  106~(11) (2022) 114040.
\newblock \href {http://arxiv.org/abs/2208.00355} {\path{arXiv:2208.00355}},
  \href {https://doi.org/10.1103/PhysRevD.106.114040}
  {\path{doi:10.1103/PhysRevD.106.114040}}.

\bibitem{Hu:2022ctr}
Z.~Hu, S.~Xu, C.~Mondal, X.~Zhao, J.~P. Vary, {Transverse momentum structure of
  proton within the basis light-front quantization framework}, Phys. Lett. B
  833 (2022) 137360.
\newblock \href {http://arxiv.org/abs/2205.04714} {\path{arXiv:2205.04714}},
  \href {https://doi.org/10.1016/j.physletb.2022.137360}
  {\path{doi:10.1016/j.physletb.2022.137360}}.

\bibitem{Collins:1981va}
J.~C. Collins, D.~E. Soper, {Back-To-Back Jets: Fourier Transform from B to
  K-Transverse}, Nucl. Phys. B 197 (1982) 446--476.
\newblock \href {https://doi.org/10.1016/0550-3213(82)90453-9}
  {\path{doi:10.1016/0550-3213(82)90453-9}}.

\bibitem{Catani:2000vq}
S.~Catani, D.~de~Florian, M.~Grazzini, {Universality of nonleading logarithmic
  contributions in transverse momentum distributions}, Nucl. Phys. B 596 (2001)
  299--312.
\newblock \href {http://arxiv.org/abs/hep-ph/0008184}
  {\path{arXiv:hep-ph/0008184}}, \href
  {https://doi.org/10.1016/S0550-3213(00)00617-9}
  {\path{doi:10.1016/S0550-3213(00)00617-9}}.

\bibitem{Bozzi:2005wk}
G.~Bozzi, S.~Catani, D.~de~Florian, M.~Grazzini, {Transverse-momentum
  resummation and the spectrum of the Higgs boson at the LHC}, Nucl. Phys. B
  737 (2006) 73--120.
\newblock \href {http://arxiv.org/abs/hep-ph/0508068}
  {\path{arXiv:hep-ph/0508068}}, \href
  {https://doi.org/10.1016/j.nuclphysb.2005.12.022}
  {\path{doi:10.1016/j.nuclphysb.2005.12.022}}.

\bibitem{Bozzi:2008bb}
G.~Bozzi, S.~Catani, G.~Ferrera, D.~de~Florian, M.~Grazzini,
  {Transverse-momentum resummation: A Perturbative study of Z production at the
  Tevatron}, Nucl. Phys. B 815 (2009) 174--197.
\newblock \href {http://arxiv.org/abs/0812.2862} {\path{arXiv:0812.2862}},
  \href {https://doi.org/10.1016/j.nuclphysb.2009.02.014}
  {\path{doi:10.1016/j.nuclphysb.2009.02.014}}.

\bibitem{Brodsky:1997de}
S.~J. Brodsky, H.-C. Pauli, S.~S. Pinsky, {Quantum chromodynamics and other
  field theories on the light cone}, Phys. Rept. 301 (1998) 299--486.
\newblock \href {http://arxiv.org/abs/hep-ph/9705477}
  {\path{arXiv:hep-ph/9705477}}, \href
  {https://doi.org/10.1016/S0370-1573(97)00089-6}
  {\path{doi:10.1016/S0370-1573(97)00089-6}}.

\bibitem{Glazek:1992aq}
S.~D. Glazek, R.~J. Perry, {Special example of relativistic Hamiltonian field
  theory}, Phys. Rev. D 45 (1992) 3740--3754.
\newblock \href {https://doi.org/10.1103/PhysRevD.45.3740}
  {\path{doi:10.1103/PhysRevD.45.3740}}.

\bibitem{Brodsky:2014yha}
S.~J. Brodsky, G.~F. de~Teramond, H.~G. Dosch, J.~Erlich, {Light-Front
  Holographic QCD and Emerging Confinement}, Phys. Rept. 584 (2015) 1--105.
\newblock \href {http://arxiv.org/abs/1407.8131} {\path{arXiv:1407.8131}},
  \href {https://doi.org/10.1016/j.physrep.2015.05.001}
  {\path{doi:10.1016/j.physrep.2015.05.001}}.

\bibitem{Meissner:2008ay}
S.~Meissner, A.~Metz, M.~Schlegel, K.~Goeke, {Generalized parton correlation
  functions for a spin-0 hadron}, JHEP 08 (2008) 038.
\newblock \href {http://arxiv.org/abs/0805.3165} {\path{arXiv:0805.3165}},
  \href {https://doi.org/10.1088/1126-6708/2008/08/038}
  {\path{doi:10.1088/1126-6708/2008/08/038}}.

\bibitem{Collins:1992kk}
J.~C. Collins, {Fragmentation of transversely polarized quarks probed in
  transverse momentum distributions}, Nucl. Phys. B 396 (1993) 161--182.
\newblock \href {http://arxiv.org/abs/hep-ph/9208213}
  {\path{arXiv:hep-ph/9208213}}, \href
  {https://doi.org/10.1016/0550-3213(93)90262-N}
  {\path{doi:10.1016/0550-3213(93)90262-N}}.

\bibitem{Bacchetta:2020vty}
A.~Bacchetta, F.~G. Celiberto, M.~Radici, P.~Taels,
  {Transverse-momentum-dependent gluon distribution functions in a spectator
  model}, Eur. Phys. J. C 80~(8) (2020) 733.
\newblock \href {http://arxiv.org/abs/2005.02288} {\path{arXiv:2005.02288}},
  \href {https://doi.org/10.1140/epjc/s10052-020-8327-6}
  {\path{doi:10.1140/epjc/s10052-020-8327-6}}.

\bibitem{Kogut:1969xa}
J.~B. Kogut, D.~E. Soper, {Quantum Electrodynamics in the Infinite Momentum
  Frame}, Phys. Rev. D 1 (1970) 2901--2913.
\newblock \href {https://doi.org/10.1103/PhysRevD.1.2901}
  {\path{doi:10.1103/PhysRevD.1.2901}}.

\bibitem{Schwartz:2014sze}
M.~D. Schwartz, {Quantum Field Theory and the Standard Model}, Cambridge
  University Press, 2014.

\bibitem{Lepage:1980fj}
G.~P. Lepage, S.~J. Brodsky, {Exclusive Processes in Perturbative Quantum
  Chromodynamics}, Phys. Rev. D 22 (1980) 2157.
\newblock \href {https://doi.org/10.1103/PhysRevD.22.2157}
  {\path{doi:10.1103/PhysRevD.22.2157}}.

\bibitem{Bacchetta:2019qkv}
A.~Bacchetta, G.~Bozzi, M.~G. Echevarria, C.~Pisano, A.~Prokudin, M.~Radici,
  {Azimuthal asymmetries in unpolarized SIDIS and Drell-Yan processes: a case
  study towards TMD factorization at subleading twist}, Phys. Lett. B 797
  (2019) 134850.
\newblock \href {http://arxiv.org/abs/1906.07037} {\path{arXiv:1906.07037}},
  \href {https://doi.org/10.1016/j.physletb.2019.134850}
  {\path{doi:10.1016/j.physletb.2019.134850}}.

\bibitem{Rodini:2022wki}
S.~Rodini, A.~Vladimirov, {Definition and evolution of transverse momentum
  dependent distribution of twist-three}, JHEP 08 (2022) 031, [Erratum: JHEP
  12, 048 (2022)].
\newblock \href {http://arxiv.org/abs/2204.03856} {\path{arXiv:2204.03856}},
  \href {https://doi.org/10.1007/JHEP08(2022)031}
  {\path{doi:10.1007/JHEP08(2022)031}}.

\bibitem{Chavez:2021koz}
J.~M.~M. Ch\'avez, V.~Bertone, F.~De~Soto~Borrero, M.~Defurne, C.~Mezrag,
  H.~Moutarde, J.~Rodr\'\i{}guez-Quintero, J.~Segovia, {Accessing the Pion 3D
  Structure at US and China Electron-Ion Colliders}, Phys. Rev. Lett. 128~(20)
  (2022) 202501.
\newblock \href {http://arxiv.org/abs/2110.09462} {\path{arXiv:2110.09462}},
  \href {https://doi.org/10.1103/PhysRevLett.128.202501}
  {\path{doi:10.1103/PhysRevLett.128.202501}}.

\end{thebibliography}
\end{document}